\documentclass[sigconf,10pt,preprint,colorinlistoftodos]{acmart}

\usepackage{xspace} 
\usepackage{enumitem}
\usepackage{color}
\usepackage{colortbl}
\usepackage{xcolor}
\usepackage{pifont}

\usepackage{epstopdf}
\usepackage{graphicx}
\usepackage{subcaption}
\usepackage{pgfplots}
\usepackage{pgfplotstable}

\usepackage[capitalise]{cleveref}

\usepackage{siunitx}
\usepackage{ifthen}

\usepackage{listings}
\usepackage[utf8]{inputenc}

\usepackage{booktabs}  
\usepackage{tabu}
\usepackage{array}
\usepackage{multirow}
\usepackage{diagbox}

\usepackage{balance}
\usepackage{microtype}

\usepackage[english]{babel}
\usepackage{blindtext}

\def\setuppreprint{1}
\def\setupcameready{0}

\newcommand{\name}{nanoPU\xspace}

\newcommand{\titleinfo}{Enabling the Reflex Plane with the nanoPU} 
\newcommand{\authorinfo}{\vspace{-20pt} Stephen Ibanez, Alex Mallery, Serhat Arslan, Theo Jepsen, Muhammad Shahbaz$^{\dagger}$, Changhoon Kim, and Nick McKeown}
\newcommand{\institutioninfo}{Stanford University~~~$^{\dagger}$Purdue University}
\newcommand{\submissioninfo}{Submission \#1}

\if\setupcameready1
    \def\shownames{1} 
    \def\showccs{1} 
    \def\showcomments{0} 
    \def\showacks{1} 

    \setcopyright{acmcopyright}
    \copyrightyear{2018}
    \acmYear{2018}
    \acmDOI{10.1145/1122445.1122456}
    
    \acmConference[Woodstock '18]{Woodstock '18: ACM Symposium on Neural
        Gaze Detection}{June 03--05, 2018}{Woodstock, NY}
    \acmBooktitle{Woodstock '18: ACM Symposium on Neural Gaze Detection,
        June 03--05, 2018, Woodstock, NY}
    \acmPrice{15.00}
    \acmISBN{978-1-4503-9999-9/18/06}
    \acmSubmissionID{123-A56-BU3}
    
    \def\removepageheaders{0} 
\else
    \def\shownames{1}
    \def\showccs{0}
    
    \def\showacks{0}
    \if\setuppreprint1
        \def\showcomments{0}
    \else
        \def\showcomments{1}
    \fi

    \setcopyright{none}
    \acmDOI{}
    
    \acmISBN{}
    
    \renewcommand\footnotetextcopyrightpermission[1]{} 
    
    \def\removepageheaders{1}
\fi

\newcommand{\ie}{i.e.}
\newcommand{\eg}{e.g.}

\newcommand{\shahbaz}[1]{\todo[color=blue]{Shahbaz: #1}}

\lstdefinelanguage{RISCV}
{
  morekeywords={
  	csrwi, csrr, bnez, beqz, mv, addi, j
  },
  morecomment=[l]{//},
  sensitive=true,
  breaklines=true,
  frame=l,
  captionpos=b,
  extendedchars=true,
  belowcaptionskip=1\baselineskip,
  xleftmargin=\parindent,
  basicstyle=\footnotesize\ttfamily,
  commentstyle=\itshape\color{green!40!black},
  numbers=left,
  numberstyle=\scriptsize\color{blue}\texttt,
  numbersep=5pt
}

\if\showcomments1
    \usepackage[textsize=tiny]{todonotes}
\else
    \usepackage[disable]{todonotes}
\fi

\begin{document}

\title{\titleinfo}

\if\shownames1
    \author{\authorinfo}
    \affiliation{
        \institution{\em \institutioninfo}
        \country{}
    }
\else
    \author{\submissioninfo}
    \affiliation{
        \institution{\pageref{lastpage} Pages Body, \pageref{totalpage} Pages Total}
        \country{}
    }
\fi

\if\showcomments1
    \setcounter{page}{0}
    \listoftodos{}
    \clearpage
\fi

\begin{abstract}
Many recent papers have demonstrated fast in-network computation using programmable switches, running many orders of magnitude faster than CPUs. 
The main limitation of writing software for switches is the constrained programming model and limited state. 
In this paper we explore whether a new type of CPU, called the nanoPU, offers a useful middle ground, with a familiar C/C++ programming model, and potentially many terabits/second of packet processing on a single chip, with an RPC response time less than \SI{1}{$\mu$}s. 
To evaluate the nanoPU, we prototype and benchmark three common network services: packet classification, network telemetry report processing, and consensus protocols on the nanoPU. 
Each service is evaluated using cycle-accurate simulations on FPGAs in AWS. 
We found that packets are classified 2$\times$ faster and INT reports are processed more than an order of magnitude quickly than state-of-the-art approaches. 
Our production quality Raft consensus protocol, running on the nanoPU, writes to a 3-way replicated key-value store (MICA) in \SI{3}{$\mu$}s, twice as fast as the state-of-the-art, with 99\% tail latency of only \SI{3.26}{$\mu$}s.

To understand how these services can be combined, we study the design and performance of a {\em network reflex plane}, designed to process telemetry data, make fast control decisions, and update consistent, replicated state within a few microseconds. 
\end{abstract}

\if\showccs1


\fi


\maketitle

\if\removepageheaders1
    \pagestyle{plain}
\fi

\begin{sloppypar}

\section{Introduction}
\label{sec:intro}
In recent years, the networking industry, research, and open-source communities have taken a special interest in {\em in-network computation}, in which computation is accelerated by programmable network devices, such as programmable switches and FPGAs~\cite{netchain,silkroad,netcache,netpaxos}. 
Sometimes the choice is clear: programmable switches can process packets at several orders of magnitude higher throughput than CPUs for about the same power consumption (\eg, a \SI{12.8}{Tb/s} switch processes about 500 times more packets per second than a CPU, yet consumes a similar amount of power), and FPGAs are about an order of magnitude slower than switches. 
Hence, in-network applications (such key-value caches~\cite{netcache}, consensus protocols~\cite{netpaxos,netchain} and L4 load-balancing~\cite{silkroad}) can run hundreds or even thousands of times faster on a switch than a CPU. 
The trade-off is also clear: Programmable switches and FPGAs are much harder to program than CPUs, with limited resources, constrained programming models, and programming languages (\eg, P4 and Verilog) that are unfamiliar to most application developers. 
Generally speaking, an in-network accelerator makes sense if, and only if, it offers at least an order of magnitude improved performance, or if it is a much cheaper or lower power solution than a conventional CPU.

With a seemingly never-ending demand for massive-scale applications requiring high throughput, low-latency reliable communications, it is worth asking how we could make it easier for developers to offload some functions into the network, particularly services that are of general utility to many applications. 
For example, Raft~\cite{raft} is a well-known consensus algorithm that allows a set of servers to agree, and perform computations, upon a set of distributed values, even when some of the servers fail. 
Because consensus protocols are hard to get right, it is worth having a standard trusted set of libraries available for developers to use and build upon. 
And, because consensus protocols are complicated, they can easily be the bottleneck for large distributed applications. 
It is therefore worth considering a highly optimized, in-network service usable by many developers of, say, a cloud provider. 
NetPaxos was a good attempt to do this for the Paxos algorithm, using P4-programmable switches, and was shown to reduce the decision latency from \SI{1.39}{ms} on a CPU to \SI{0.37}{ms} on a switch~\cite{netpaxos}. 

Another example of an in-network service is processing network telemetry data. 
With increased interest in observing what the network is doing, for example to monitor SLOs and to improve security, telemetry mechanisms (such as INT~\cite{INT}) are being increasingly deployed. 
In one approach, every packet carries telemetry data collected from switches along its path, including switch ID, queueing delay, and buffer occupancy. 
Researchers have demonstrated how INT data can detect and diagnose microbursts~\cite{conquest}, enable congestion-aware routing~\cite{hula} and assist congestion control~\cite{hpcc}. 
One challenge when deploying INT is the ``firehose'' of measurement data carried by packets, requiring significant processing of telemetry data to build a complete picture of the network state~\cite{sonata, intcollector}. 
For example, a \SI{100}{GB} link carrying \SI{1500}{B} messages can generate 8.3 million telemetry reports per second, way beyond the abilities of a regular CPU to process. 
A programmable switch or FPGA could pre-process telemetry reports, to remove duplicates, build flow reports, and implement real-time thresholds and alarms. 
But these platforms have limited state and are harder to program than a CPU.  
Telemetry would be easier to deploy if we could provide easy to program, cost-effective, in-network elements to process INT data.

Therefore, the goal of this paper is to explore and evaluate whether the recently proposed {\em nanoPU}~\cite{nanopu}---a high throughput, low latency CPU optimized for packet processing---would make an effective, efficient, high performance platform for a broad class of in-network services. 

We originally proposed the nanoPU for very low-latency RPCs lasting less than \SI{1}{$\mu$}s, which we call {\em nanoServices}. 
To the programmer, the nanoPU is a regular multicore CPU, with an extremely fast, low-latency network interface. 
Arriving packet data is placed directly into a dedicated CPU register, in less than \SI{40}{ns} of arriving from Ethernet, or about an order of magnitude faster than the conventional path through PCIe, memory and cache hierarchy. 
A nanoPU core can send and receive at sustained \SI{200}{Gb/s}, and a nanoPU device could contain many such cores (and interfaces). 
In essence, the nanoPU treats packet data as a first-class citizen, just as a RISC CPU core is optimized for load-store memory operations. 
The nanoPU provides two separate paths, one direct through the register and one through memory (Section~\ref{sec:background}).

The nanoPU is designed to combine the ease and familiarity of programming a CPU with a blazingly fast network interface. 
A nanoPU chip could be built today containing over 500 RISC-V cores running at \SI{3.2}{GHz} with 128 network interfaces running at \SI{100}{Gb/s}, processing over 10 billion packets per second at a sustained throughput of \SI{12.8}{Tb/s}, all in one package with the same size and power as a modern switch ASIC. 
If the data needed to handle a short RPC request is cache-resident, each core could respond to an RPC request in less than \SI{100}{ns}. 
The nanoPU achieves this by placing several key functions in hardware: reliable transport~\cite{nanotransport} (\ie, can be programmed to run protocols such as NDP~\cite{ndp}), core selection, and thread scheduling. 
Such a device could greatly accelerate large applications that are able to shard data and computation across many nanoPU cores.

In this paper, we explore how---in addition to providing very low-latency RPCs---the nanoPU can be used to accelerate in-network services. 
Specifically, we evaluate three common network services: \emph{packet classification}, \emph{network telemetry report processing}, and \emph{the Raft consensus protocol}.
These services are commonly used as building blocks for many distributed applications and network control systems.
We implement each of these services on our simulated nanoPU prototype running on FPGA F1 instances in AWS.
We find that a production grade implementation of the Raft consensus protocol running on the nanoPU runs 2$\times$ faster than the state-of-the-art. 
Of particular interest is the nanoPU's support for latency-bounded services: Our 3-way replicated key-value store, built on Raft, completes on average in \SI{3}{$\mu$}s with a 99\% tail latency of just \SI{3.26}{$\mu$}s. 
Similarly, our nanoPU prototype improves throughput and reduces latency of packet classification and network telemetry report processing by a factor of 2 and 20, respectively, relative to the state-of-the-art running on a conventional CPU.

To further motivate the utility of these performance gains, we describe how the three network services can be composed together to enable support for real-time network control, something we call the {\em reflex plane}.
Real-time network control is very challenging in modern networks because (1) forwarding plane devices, such as switches, are unable to perform sophisticated packet processing, and (2) the control plane running on general-purpose CPUs is too slow to utilize fine-grained telemetry measurements.
The nanoPU is well positioned to implement the reflex plane because it works faster than today's control-plane platforms (that run on CPUs) and yet is more flexible and easier to program than today's data-plane platforms (running on switch ASICs).

This paper makes the following contributions:
\begin{itemize}[leftmargin=*]
    \item Provides open-source implementations of three important network services running on the nanoPU (Section~\ref{sec:implementation}): packet classification, telemetry report processing, and RAFT consensus. This is in addition to the open-source implementation of the nanoPU itself, in the Chisel language, running on the Firesim platform~\cite{firesim}.
    \item Demonstrates significant performance improvements for those services (Section~\ref{sec:eval}).
    \item Describes how those services can be composed to enable real-time closed-loop control of a network (Section~\ref{sec:reflex-plane}).
\end{itemize}
\section{Background: The nanoPU}
\label{sec:background}

\begin{figure}
  \includegraphics[width=1\linewidth]{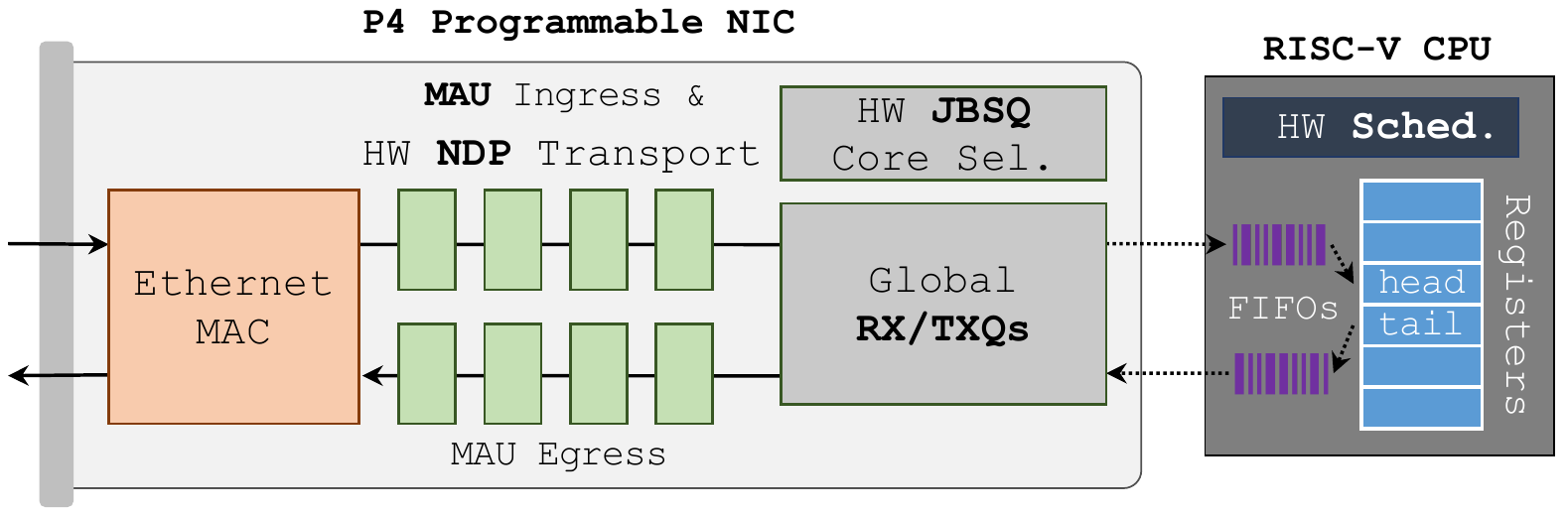}
  \vspace{-20pt}
  \caption{The \name{} prototype. Total wire-to-wire latency is {\bf 65ns}.}
  \label{fig:prototype}
\end{figure}

\shahbaz{update Figure 1}

The nanoPU~\cite{nanopu} was recently proposed as a new class of CPU optimized for processing RPC requests in less than \SI{1}{$\mu$}s.
The nanoPU borrows the low-latency Lightning NIC design~\cite{lnic} in order to minimize the tail latency of RPC requests (\Cref{fig:prototype}).
Specifically, it addresses the following three issues that contribute to high RPC tail latency:

\begin{itemize}[leftmargin=*]
    \item \textbf{Memory and cache hierarchy (on the critical path).} The networking subsystem of today's end hosts uses memory as a workspace to hold and process packets. This inherently leads to interference with applications' memory accesses, introducing resource contention, which causes poor RPC tail latency. Furthermore, if a packet is transferred over PCIe to DRAM, it is not available to the CPU until several hundred nanoseconds after it arrives~\cite{ramcloud}. While direct cache access technologies reduce such latency, the packet must still go through many layers of the networking stack.
    \item \textbf{Suboptimal scheduling.} Various scheduling schemes are involved in RPC handling to dispatch arriving packets to cores for network stack processing and to relegate RPC requests to cores for response generation. At each step, a software {core-selection algorithm} selects the core, and a {thread scheduler} decides when processing begins. Both algorithms require frequent access to memory by the cores and the NIC, requiring mediation of the memory bus, PCIe, and cache lines. 
    \item \textbf{Network congestion.} While network congestion can significantly increase RPC latency for obvious reasons, the way today's software-based transport protocol implementations process packets to manage and recover from congestion also gets in the way of reducing RPC latency.
\end{itemize}

Current systems attempt to tackle a subset of these problems, but no existing system addresses all three simultaneously.
Nebula~\cite{nebula} attempts to address the problem of memory bandwidth interference and implements an efficient core-selection algorithm in hardware, but its proposed approach is less effective when RPC processing time is unknown or highly variable.
Shinjuku~\cite{shinjuku} and ZygOS~\cite{zygos} are low-latency operating systems that aim to efficiently load-balance and schedule request processing across cores. 
As software solutions, however, they are unable to operate efficiently at a granularity below \SI{5}{$\mu$}s making them ill-suited for sub-microsecond RPCs.
eRPC~\cite{eRPC} achieves impressive low-latency results on commodity hardware---approaching that of hardware accelerators---by optimizing for the common case. 
However, these common-case optimizations sacrifice tail latency.
Lastly, RDMA gives applications direct and low-latency access to a remote server's memory~\cite{herd, fasst, drtmr}.
However, the \name{} targets RPC-based applications that need low-latency access to remote CPUs, {\em not} remote memory.

\begin{figure}[t]
  \centering
  \footnotesize
  \begin{minipage}[c]{0.9\linewidth}
\begin{lstlisting}[language=riscv]
// Simple loopback & increment application
entry:
  // Register port number & priority with NIC
  csrwi	lcurport, 0
  csrwi	lcurpriority, 0
  csrwi	lniccmd, 1

// Wait for a message to arrive
wait_msg:
  csrr	a5, lmsgsrdy
  bnez	a5, loopback_plus1_16B
idle:
  csrwi	lidle, 1 // app is idle
  csrr	a5, lmsgsrdy
  beqz	a5, idle

// Loopback and increment 16B message
loopback_plus1_16B:
  mv netTX, netRX // copy app hdr from rx to tx
  addi netTX, netRX, 1 // send word one + 1
  addi netTX, netRX, 1 // send word two + 1
  csrwi	lmsgdone, 1 // msg processing complete
  j wait_msg // wait for the next message
\end{lstlisting}
  \end{minipage}
  \vspace{-5pt}
  \caption{Loopback with increment. A nanoPU RISC-V assembly program that waits for a 16B message, increments each word, and returns it to the sender.}
  \label{lst:asm}
\end{figure}
 
As shown in Figure~\ref{fig:prototype}, the nanoPU design includes a number of features that differentiate it from existing approaches.
First and foremost, the nanoPU tightly integrates a \SI{200}{Gb/s} NIC with RISC-V Rocket cores~\cite{rocket-chip} and uses dedicated two-level FIFOs (i.e., global and local RX \& TX queues) for network packets, placing RPC requests directly into the CPU register file, eliminating the need for network data to traverse a PCIe bus or the CPU memory hierarchy.
The wire-to-wire latency for a loopback application running on the nanoPU is just \SI{65}{ns}, which is 13$\times$ faster than the state-of-the-art.
Each core is able to saturate a \SI{200}{Gb/s} network link if it dedicates all instructions to performing network IO, 2.5$\times$ faster than the state-of-the-art.
The nanoPU NIC implements NDP transport~\cite{ndp} in hardware, to ensure low tail latency through the network fabric by keeping queues small, even under challenging incast workloads.
The nanoPU NIC also implements JBSQ~\cite{r2p2, nebula} core-selection in hardware to evenly distribute network load across CPU cores.
The \SI{200}{Gb/s} NIC can process more than 350 Million packets/second.
The nanoPU offloads thread scheduling decisions to hardware, enabling 99\% tail latencies below \SI{2.1}{$\mu$}s under high load for high-priority applications, even when the core is shared with low-priority background tasks.

We believe that the nanoPU's extremely low-latency network stack, combined with its high per-core throughput, make it an ideal platform for accelerating a broad class of in-network services.
The following section demonstrates the basic mechanics of how software running on the nanoPU interacts with the hardware.

\subsection{The Hardware/Software Interface}
Listing~\ref{lst:asm} shows a simple loopback-and-increment program in RISC-V assembly running on the nanoPU.
The program continuously reads 16B messages (two 8B integers) from the network, increments the integers, and sends messages back to their sender.
The program details are described below.

The \verb|entry| procedure binds the thread to a layer-4 port number at the given priority level by first writing a value to both the \verb|lcurport| and \verb|lcurpriority| control and status registers (CSRs), then writing the value 1 to the \verb|lniccmd| CSR.
The \verb|lniccmd| CSR is a bit vector used by software to send commands to the networking hardware; in this case, it is used to tell the hardware to allocate both RX and TX queues for port 0 at priority 0.
The \verb|lniccmd| CSR can also be used to unbind a port or to update the priority level.

The \verb|wait_msg| procedure waits for a message to arrive in the RX queue by polling the \verb|lmsgsrdy| CSR until it is set by the hardware.
While it is waiting, the application tells the hardware thread scheduler that it is idle by writing to the \verb|lidle| CSR during the polling loop.
The scheduler uses the idle signal to evict idle threads in order to schedule a new thread that has messages waiting to be processed.

The \verb|loopback_plus1_16B| procedure simply swaps the source and destination addresses by moving the RX application header (the first word of every received message, which indicates the source IP/port and message length) from the \verb|netRX| register to the \verb|netTX| register, shown on line 19 (Listing~\ref{lst:asm}). 
It then increments every integer in the received message and appends them to the message being transmitted.
After the procedure has finished processing the message, it tells the hardware scheduler it is done by writing to the \verb|lmsgdone| CSR.
The scheduler uses this write signal to reset the message-processing timer for the thread. 
It may also evict the thread to ensure that messages arriving for other threads of the same priority are processed in FIFO order.
Finally, the procedure waits for the next message to arrive.

Applications that use variable-length messages can use the message length (in the application header) to read the correct number of words from the network RX queue.
If an application reads an empty RX queue, the resulting behavior is undefined---similar to reading uninitialized variables.

The program described here is shown in RISC-V assembly for clarity. 
In practice, developers implement applications in C/C++.
Compiling applications to run on the nanoPU does not require any compiler modifications.
We use \verb|gcc| or \verb|g++| and pass in the appropriate flags to tell the compiler not to use the \verb|netRX| and \verb|netTX| registers for temporary storage.

\section{Fast In-Network Services}
\label{sec:motivation}

We implement and evaluate three in-network services that are commonly used as building blocks for distributed applications and network control systems: packet classification, telemetry report processing, and consensus protocols. We port each application to run on the nanoPU.

\paragraph{Packet Classification}
Packet classification involves parsing packets to extract relevant header fields, and matching them against a set of rules. 
Generalized packet classification is quite complicated~\cite{packetClassification,nuevomatch,cutsplit,tuplemerge,neurocuts}.
A rule may contain wildcard and range matches across multiple fields, and a given set of fields may match multiple rules.
Packet classification is found in virtually all packet-processing systems: switches, routers, firewalls, load-balancers, billing and accounting systems, deep-packet inspections, and so on. 

Classification is often the first operation performed by a packet-processing system, and determines how the packet is subsequently processed.
Therefore, an in-network packet classification service needs to have higher throughput to feed data into the rest of the system.
Furthermore, if the packet-processing system requires low and predictable end-to-end latency, then minimizing classification latency is crucial, as it is on the critical path for each packet.

\paragraph{Telemetry Report Processing}
\label{sec:telemetry-service}
In-band Network Telemetry (INT)~\cite{INT} is designed to provide fine-grained per-packet measurements. 
Packets carry an INT instruction telling the switches to insert measurement data into the packet header as it is forwarded along its path. 
The INT bitmap instruction tells each switch along the path to insert specific metadata values, such as the switch ID, the packet's ingress and egress ports, queue ID, queue size, hop latency, the forwarding rule(s) the packet matched upon, and current link utilization.
INT can be extended to support any metadata that is available to the switches.
An INT sink (the last switch or the end host) extracts the INT header and metadata from each packet, delivers the original data packet to the destination host network stack, and forwards the INT metadata to a monitoring system along with the relevant header fields. 

The ability to provide fine-grained (potentially per-packet) measurements is both a blessing and a curse.
On one hand, it provides an opportunity for control systems and network operators to clearly and precisely understand how the network is behaving; it provides accurate ``ground truth'' information about network conditions. 
If the INT reports are processed quickly enough, actions can potentially be taken to resolve performance, reliability or security problems. 
On the other hand, the firehose of telemetry data can easily overwhelm monitoring systems if they are not designed carefully and implemented efficiently.
In our example in the introduction, a monitoring application for a 100GE link must be able to process reports at 8.3M reports/second, almost an order of magnitude faster than state of the art single core INT processing systems can support today~\cite{intcollector}.

Modern INT processing systems are therefore not designed to process, analyze, and act upon telemetry data in real time.
Rather, they are designed to give network operators the means to retroactively diagnose issues that were observed in the past.
However, we believe that the true power of fine grained telemetry lies in the opportunity to diagnose and mitigate issues in real-time, allowing for closed-loop control.
Achieving this goal will require systems that not only support high report processing throughput, but also allow for extremely low response times in order to quickly react to observed issues. 
In modern networks, these issues may last a few RTTs (tens of microseconds), yet cause trouble for highly distributed, tail latency-sensitive applications.

\paragraph{Consensus Protocols}
Consensus protocols enable replicated state machines to serve as a fundamental building block for many distributed applications, including replicated in-memory key-value stores~\cite{mica}, SDN controllers~\cite{onos}, distributed process coordinators~\cite{zookeeper}, and lock management systems~\cite{chubby}.

A consensus protocol allows a set of distributed nodes to agree upon shared state values, even if some of the nodes fail. 
The performance of consensus protocols is often determined by the communication latency between remote nodes, particularly when state update operations are simple. 
Network latency is therefore a key performance metric for consensus protocols. Furthermore, since an operation is not considered complete until the last replica indicates its completion, minimizing tail latency is of utmost importance.

\subsection{The nanoPU Service Prototypes}
\label{sec:implementation}

We implement and evaluate the three network services described in Section~\ref{sec:motivation}: packet classification, telemetry report processing, and consensus protocols. 
Each service is ported to run on the nanoPU prototype, running on the cycle-accurate simulator in AWS F1 FPGA instances and managed by the Firesim framework~\cite{firesim}.

\subsubsection{Multi-field Packet Classification}
Programmable network switches, such as Tofino~\cite{tofino}, contain hardware support for programmable parsing (implemented using a state machine) and packet classification (implemented by associative matches in TCAMs and SRAMs) at line rate. 
Similarly, the nanoPU NIC contains a P4-programmable PISA pipeline that can classify packets at line rate using TCAMs and SRAMs in the match-action units of each stage. 
For example, our \SI{200}{Gb/s} NIC can perform associative lookups and hence can classify 350 million packets/second for, say, an access-control list (ACL) lookup.  

However, while fast, the lookup tables in a NIC or a switch are necessarily quite small (\eg, the nanoPU NIC prototype has space for only about 10K ACLs~\cite{xilinx-tcam}). 
This is insufficient for many in-network services, such as billing systems, QoS routers, or an NFV intrusion detection system looking for matches in a 100K rule set~\cite{classbench}.

Many authors have instead proposed {\em software}-based classification algorithms, designed to work with much larger rule sets~\cite{packetClassification}. 
In our case, because the nanoPU runs fastest when data is cache-resident, we can partition the rule set over multiple nanoPU cores, then use the P4 pipeline to load-balance incoming packets to one or more cores, by calculating a hash of the packet header. 

Our prototype uses the recently proposed NuevoMatch~\cite{nuevomatch} multi-field packet classification algorithm, currently the fastest software algorithm we are aware of. 
NuevoMatch uses a novel data structure (RQ-RMI) to replace memory queries (normally the bottleneck for software classification) with model inference computations that are guaranteed to be correct. 
RQ-RMI rules are compressed into model weights that fit into the on-die CPU cache.  
The authors demonstrate that using 500K multi-field rules from the standard ClassBench benchmark, lookups are at least 1.6$\times$ faster, and rules compressed at least 4.9$\times$ more than the best state of the art algorithms~\cite{cutsplit,neurocuts,tuplemerge}.  

Our prototype runs the NuevoMatch algorithm on the nanoPU. 
Because multi-field packet classification is stateless (\ie, no state is recorded and shared between packets) and the rule set is read-only (except for occasional changes written by a control program), we can safely replicate the rule set across nanoPU cores or partition it into non-overlapping regions of header space~\cite{HSA}. 
The nanoPU can load-balance across cores two different ways: First, the NIC pipeline can hash packets to a portion of the rule set running on one core. 
Second, the nanoPU contains a hardware core selection algorithm to balance packets across cores holding the same rule set. 
Thus, as long as the classification rules that a core is responsible for fit into its cache, 
we can except the nanoPU to classify packets very quickly.

Porting NuevoMatch to run on the nanoPU involved making minimal changes to the NuevoMatch library. 
NuevoMatch can batch the classification of multiple packets to take advantage of SIMD optimizations. 
In our implementation, we classify one packet at a time on each core, so we removed the SIMD batching, which lowers the latency. 
After executing the RQ-RMI model, NuevoMatch uses a remainder classifier. 
We configured NuevoMatch to use the CutSplit~\cite{cutsplit} algorithm as the remainder classifier.

\subsubsection{Path Latency Anomaly Detection}
\label{sec:report-proc-impl}

While there are many interesting per-packet monitoring applications, we chose to evaluate {\em path latency anomaly detection and mitigation}. 
The application measures the network latency encountered by packets as they traverse the network, and keeps track of the typical latency for a given flow (see below). 
If a flow experiences abnormally high network latency, the application attempts to fix (mitigate) the issue by updating switch forwarding tables to route the flow along a different path.

The application requires a monitoring node to process per-packet telemetry reports carrying the latency that the packet experienced at each hop along its path. The monitor can use many different algorithms to determine whether the flow is experiencing a short- or long-term sustained increase in latency. 
For example, it could implement a simple threshold, or compare against a moving average. 
Our prototype calculates a moving average over the last ten INT reports for the flow.
If the path latency exceeds the moving average by more than a threshold amount, the monitoring application  generates a command to tell the first switch on the path (or at the switch just before the increase) to redirect future packets along a different path.

Our goal here is not to propose this particular algorithm; rather, we pick an algorithm requiring non-negligible state and computation for each arriving INT report.
It would be difficult for a programmable switch to implement a moving average for many concurrent flows; in software, the nanoPU simply maintains a circular buffer. 
The nanoPU seems well-suited to this application because it should be able to process a very high rate of INT reports, and then respond (to re-route the flow) in less than a microsecond.

\subsubsection{Raft Consensus Protocol}
From among the many available consensus protocol implementations~\cite{netpaxos, libraft, netchain}, we choose to use Raft. 
Raft is designed to be easy to understand and is widely used in production deployments. 
We port an open-source production quality implementation of Raft~\cite{libraft} to run on the nanoPU.

In consensus protocols such as Raft, a write operation is not considered committed until the last replica completes its write, and therefore the tail latency tends to dictate the performance of the application. 
The nanoPU is optimized to minimize RPC tail latency, and has explicit support for applications wishing to strictly bound the tail latency. 
Raft should be well suited to the nanoPU.

Porting Raft to run on the nanoPU required us to make some small modifications to the library source code. 
We handled incoming network data (into the CPU register file) via an event loop that calls the Raft library callback functions. 
We also added message type IDs into the Raft data structure, adjusted the initial size of the library memory buffer, and removed some logging statements. 
Otherwise, we left the Raft source code unchanged; we include our modified version in the artifact accompanying this paper.

When adding a new state variable to its store, Raft first replicates the value, waiting for a majority of servers to signal that they have replicated it. 
Next, it `applies' the entry to each server's `state machine', where the particular state machine in use is application-dependent. 
In our case, we use the MICA key-value store~\cite{mica} with 16B keys and 64B values as our example state machine. 
Clients send the Raft cluster MICA write requests, which are replicated, committed, and then applied to the MICA databases.

\subsection{End-to-End Evaluation}
\label{sec:eval}

We evaluate our nanoPU prototypes using cycle-accurate simulations.
We use both software simulations with Verilator~\cite{verilator} and hardware-accelerated simulations running on FPGAs in AWS F1 instances using the Firesim framework~\cite{firesim}.
All of our performance evaluations are based upon a \SI{3.2}{GHz} nanoPU core clock rate.
The nanoPU prototype does not include MAC and serial IO logic.
Therefore, unless explicitly mentioned, the reported results do not include the additional latency that would be added by these modules---$\approx$\SI{26}{ns} in both the RX and TX directions.

\subsubsection{Packet Classification}
We evaluate the performance of packet classification on the nanoPU using NuevoMatch~\cite{nuevomatch}. 
We ran the same ACL1 ClassBench~\cite{classbench} workload used by NuevoMatch: 100K ACL rules and a trace of 700K packets with six 4-byte header fields. 
We use a load generator to send the trace packets to the nanoPU running on the FPGA and measure the end-to-end latency from the load generator, including \SI{43}{ns} link latency in each direction.
The nanoPU has a 16KB L1 cache and a 512KB shared last level cache (LLC).

As a baseline, we ran NuevoMatch with eRPC on a commodity x86 server with 384KB L1 cache and 15MB LLC. 
We used a testbed of two servers with Mellanox CX5 \SI{100}{Gb/s} NICs connected via a Tofino switch~\cite{tofino}. 
We used the switch to measure the end-to-end latency for the server to classify a packet and send a response.

\begin{figure}[t]
    \includegraphics[width=\linewidth]{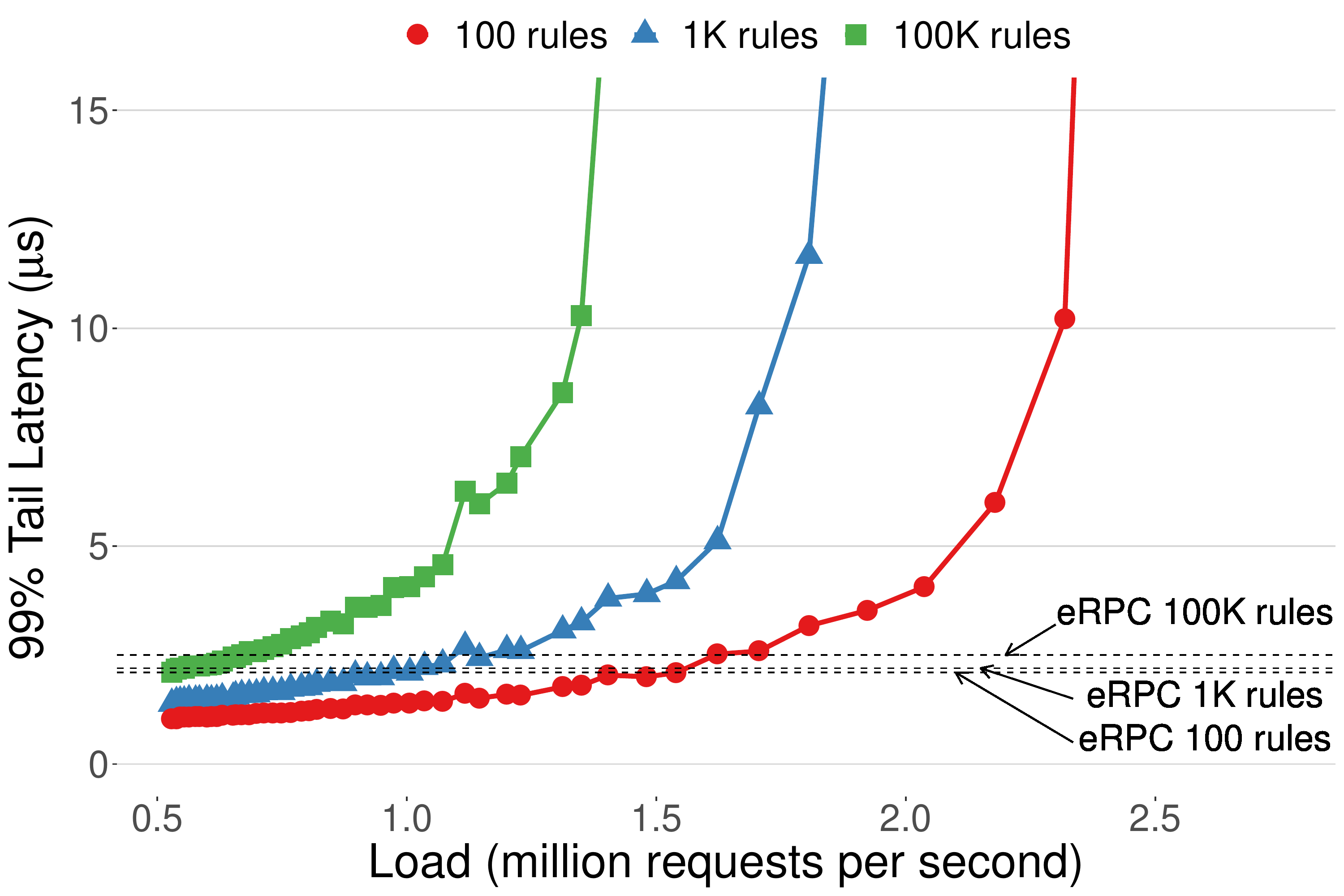}
    \vspace{-15pt}
    \caption{99\% tail latency for NuevoMatch on both the nanoPU and eRPC/x86 using various size rulesets.}
    \label{fig:class-tail-eval}
    \vspace{-10pt}
\end{figure}

Figure~\ref{fig:class-tail-eval} shows the 99\% tail latency vs load for NuevoMatch running on the nanoPU. 
It also shows the 99\% tail latency for NuevoMatch running on eRPC under low load.
While 100K rules fit in the LLC on the x86 server, they do not fit for the nanoPU.
Hence, we see that as the number of rules decrease, the benefits provided by the nanoPU increase, because the latency is less dominated by memory accesses and more by network IO, which is highly optimized on the nanoPU.
At low load with 100 rules, the 99\% tail latency for eRPC is \SI{1.9}{$\mu$}s vs \SI{1.0}{$\mu$}s when using the nanoPU.
It is reasonable to expect that future generation nanoPUs will have cache sizes that meet or exceed those on modern x86 processors.



\subsubsection{INT Report Processing}

We evaluate the performance of the path latency anomaly detection and mitigation monitoring application, as described in Section~\ref{sec:report-proc-impl}.
We measure two performance metrics: (1) maximum throughput when processing INT reports, and (2) response latency (a.k.a reflex latency) to generate a re-route command: the time from when INT telemetry metadata enters the NIC until the command to update the switch forwarding table leaves the nanoPU's NIC. 

\paragraph{Throughput:} Our first experiment measures throughput. Our goal is to process every arriving INT report for a \SI{100}{Gb/s} link, or $8.3 \mbox{M reports/second}$, assuming 1500B data packets. We send in a stream of 100 telemetry reports at line rate and then measure how long it takes the application to process all 100 reports.
The application does not generate any switch update commands, and so this experiment emulates best-case steady state behavior, when the network is running at 100\% load. 
Our results indicate that the monitoring application running on the nanoPU processes 20M reports/second using a single core.
Seeing as this is well above the target 8.3M reports/second, we can expect this application will have no problem keeping up with incoming telemetry reports.

\paragraph{Reflex Latency:} To measure response latency, we run a similar experiment to the one described above; this time, the monitoring application generates a single switch update command in response to the incoming telemetry data.
We measure a reflex response latency of \SI{78}{ns}, or about \SI{130}{ns} when including the MAC and serial IO.

\begin{table}[t]
\begin{center}
\begin{tabular}{crr}
                 & \textbf{Throughput (Mrps)} & \textbf{Reflex Latency (ns)} \\ \toprule
{nanoPU}  & 20                         & 130               \\ \hline
{eRPC}    & 0.7                        & 2400                
\end{tabular}
\caption{Comparison of telemetry report processing throughput and reflex latency for both the nanoPU and eRPC.}
\label{tab:int-results}
\vspace{-10pt}
\end{center}
\end{table}

Table~\ref{tab:int-results} compares the per-core throughput and average reflex latency of our nanoPU prototype against the same monitoring application using eRPC on a traditional x86 host CPU with a Mellanox CX5 \SI{100}{Gb/s} NIC. 
We measure the reflex latency using hardware timestamping on a Tofino switch and account for the link latency between the switch and host.
We see that the nanoPU improves throughput and reflex latency by about 30$\times$ and 20$\times$, respectively.
On both the nanoPU and the x86 server, the time to process each report is about \SI{50}{ns}.
However, eRPC introduces additional software overheads (e.g. congestion control) which dramatically reduce the throughput of the system.
On the nanoPU, the application logic is the only thing running in software, everything else is implemented in pipelined logic in hardware.
This allows the system to sustain significantly higher throughput than modern systems, especially for applications with very short processing times, like this one.
The nanoPU also improves reflex latency as a result of its highly optimized network stack which is directly connected to the CPU register file.

\subsubsection{Raft Consensus Protocol}
We evaluate Raft by running a 16B-key, 64B-value MICA key-value store state machine. 
For our experiment, we use four simulated nanoPUs (each running one core). 
Three nanoPUs form the Raft cluster, and the fourth serves as the Raft client. 
A single simulated switch connects the nanoPU client and the three nanoPU Raft cluster. 
The links between each nanoPU and the central switch are configured with a \SI{43}{ns} latency (about 12m cables), and the switch itself forwards with a simulated latency of either \SI{300}{ns} (equal to low latency cut-through commercial switch ASICs~\cite{SX1036}) or \SI{1}{ns} (to benchmark the non-switch latency). 
Although our Raft cluster correctly implements leader election and can tolerate server failure, and our client can automatically identify a new Raft leader, we conduct performance tests of a Raft cluster in the steady-state, failure-free case, with a single steady leader and three fully functioning replicas.

We measure the latency from when the client issues a three-way replicated write request to the Raft cluster, until the client hears back from the cluster leader that the request has been fully replicated and committed across all three Raft servers. 
In 10,000 trials, with a switch latency of \SI{1}{ns}, we measured a \SI{1.88}{$\mu$}s median and \SI{2.06}{$\mu$}s 99\% tail latency under no load at the nanoPU client. 
With a more realistic \SI{300}{ns} switch latency, the average latency would therefore be \SI{3.08}{$\mu$}s, with a \SI{3.26}{$\mu$}s 99\% tail latency. eRPC~\cite{eRPC}, a high performance, highly optimized RPC library for general purpose CPUs reports a \SI{5.5}{$\mu$}s median and \SI{6.3}{$\mu$}s 99\% tail latency --- about a factor of 2 higher latency than the nanoPU implementation.

\begin{figure}
    \includegraphics[width=\linewidth]{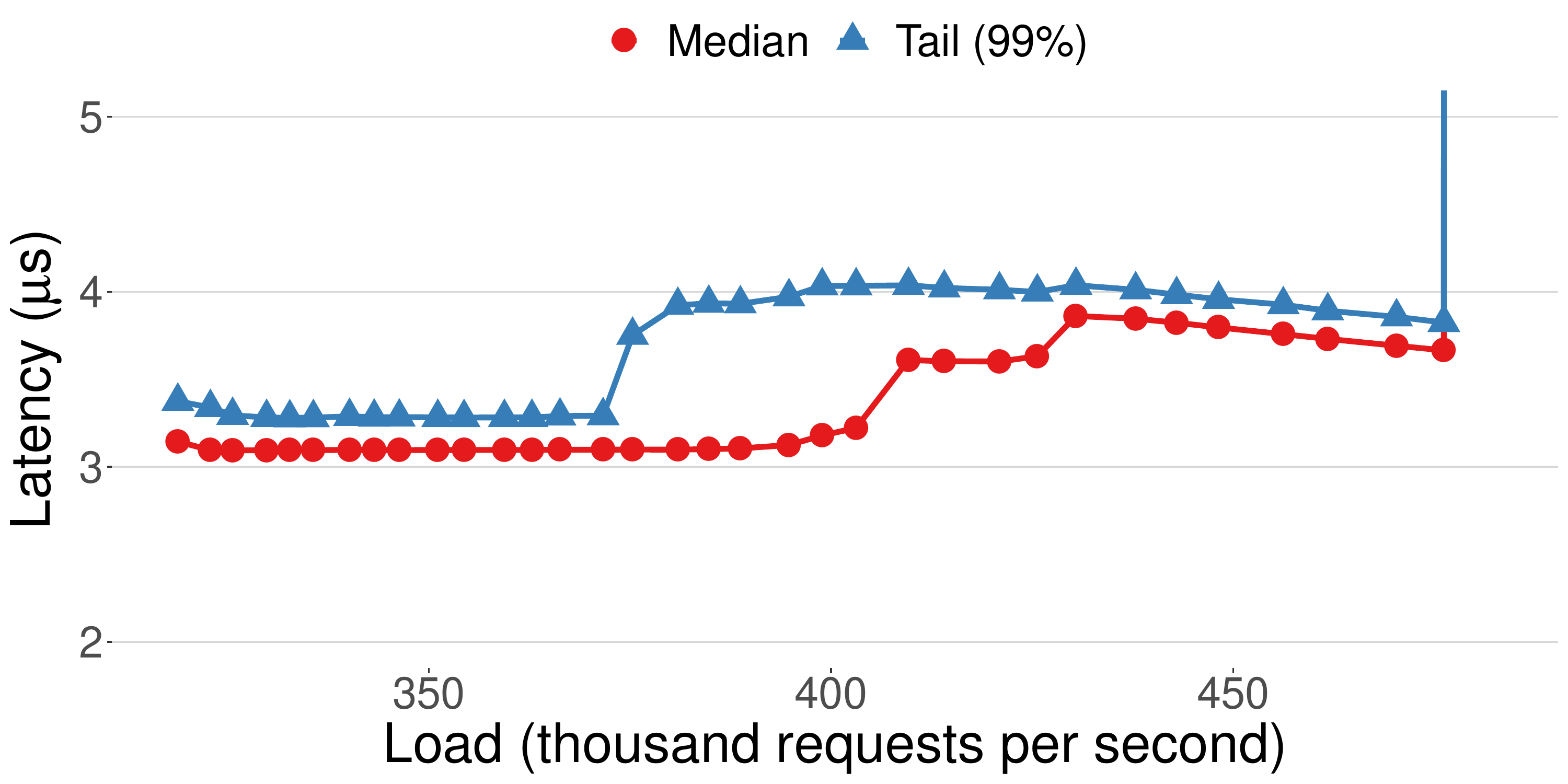}
    \caption{Median and 99\% tail latency for Raft three-way replicated writes, as measured by the client, as a function of the load.}
    \label{fig:raft_tail_eval}
    \vspace{10pt}
\end{figure}

We also evaluate a nanoPU Raft cluster's 99\% tail latency across a range of loads, beginning at 318,000 requests per second and continuing to 507,000 requests per second. 
The results are shown in Figure~\ref{fig:raft_tail_eval}. As can be seen in the figure, the nanoPU maintains a low 99\% tail latency (3--\SI{4}{$\mu$}s) up to approximately 500,000 requests per second, at which point the rate of incoming requests (roughly one every \SI{2}{$\mu$}s) begins to exceed the rate at which the system can process the requests, and server queues fill up causing requests to be dropped.

\section{Building a Reflex Plane}
\label{sec:reflex-plane}

Up until now, we have evaluated three common building blocks for in-network services to support large distributed applications. 
Next, we will examine one particular distributed application in much more detail, to see how these accelerated building blocks help. 
We call the system the {\em Reflex Plane}, a subsystem to help accelerate decisions made by a logically centralized SDN control plane, such as ONOS~\cite{onos} or ONIX~\cite{onix}.

This section is a design study: While everything up until this point has been prototyped and extensively benchmarked---and the means to reproduce it made available---we have not implemented the complete reflex plane system. 
Our benchmarking results from our building blocks provide evidence for how the reflex plane will perform, but it is beyond the scope of this paper to build it. 


\subsection{Motivation}
Network control planes today often use a logically centralized ``SDN'' controller that maintains a global view of the network state, allowing it to make globally optimal decisions. However, the latency of the control loop (from data plane measurement to control response) can be as high as seconds or even minutes, far from real-time control.

There are many examples of network failures, anomalies and performance issues that we would like to address via a control loop with a latency on the order of just a few network round trips (RTTs), a handful of microseconds. Unfortunately this is not practical using traditional control planes.
The problem is even more challenging for large networks, for which the control plane is a large distributed system itself with many instances, and when the network is heavily loaded. 

If the control plane is to commit to strict service level objectives (SLOs) it must react quickly to current network state in order to evenly distribute traffic, handle failures, and impose security policies. 
To make this more concrete, consider a control program whose job is to detect and mitigate microbursts, network congestion events that last for only tens or hundreds of microseconds~\cite{microburst-measurements}. 
Mitigating microbursts requires detecting them in the forwarding plane (\eg, from INT samples received by a collector), identifying the culprit flows, then responding with a re-route, source throttle, or in-network policing of just the culprit flows. 
A good decision will require access to global state and policy, and yet no control plane that we are aware of is able to respond within a few microseconds.

Recently, researchers have observed this dilemma and have proposed distributing some control functionality to programmable network switches, enabling them to react very quickly to observed issues/inconsistencies~\cite{mantis, hula, contra, poseidon}.
However, this approach has three problems: (1) Programmable switches can perform very limited functionality. 
They have limited amounts of state memory, have a limited instruction set and do not have a program counter for more complex algorithms. 
(2) A network switch does not have access to global network information and hence it is forced to make control decisions based upon only local state, which often can lead to sub-optimal decisions. 
(3) If a control plane delegates decisions to a switch, it breaks the usual ``top-down'' programming model in which the control plane tells the forwarding plane what to do; this leads to more opportunities for inconsistent state, and hence incorrect future decisions.

We therefore set the design exercise: Could we build a network control system that provides the responsiveness of the distributed approach while utilizing global information like the centralized approach? We call our solution the network reflex plane.
The reflex plane logically sits between the forwarding and control planes.
It is responsible for processing all data-plane telemetry measurements in a distributed fashion, identifying issues and inconsistencies, and providing an extremely low latency response to mitigate issues in real-time. 
It operates as a delegate of the control plane, which is still in charge. In our examples, as a general rule of thumb, the monitoring logic in the reflex plane maintains ephemeral state (\eg, recent queue occupancy information, current link utilization, current path latencies) and is not burdened with the need for large amounts of persistent global state. 
When it makes a decision, it can take certain (but not all) actions authorized by the control plane; it then must make sure its decision quickly updates part of the persistent state associated with each network element in a fault tolerant manner.

If the reflex plane is to respond in just a few microseconds, tail latency becomes king~\cite{tail-at-scale}; we must make sure decisions are both fast and predictable, even if a decision relies on many measurements. Our reflex plane programs encounter too much overhead on a conventional CPU and are too complex for a programmable switch. 
We thus study how well the reflex plane runs on a cluster of nanoPUs. 

Our reflex plane is built from the three network services evaluated earlier: (1) {\em Packet Classification} to classify telemetry reports and dispatch them to the appropriate core(s) for further processing. (2) {\em Telemetry Report Processing} to detect and respond to data plane issues quickly. (3) {\em Fault-tolerant network state management} to store the relevant state about each network element, which can be updated with very low latency operations.

\subsubsection{Example Reflex Operations}

Table~\ref{tab:reflex-examples} describes a number of operations enabled by our reflex plane.
We detail two such examples below.

\paragraph{a) Microburst Detection and Mitigation.} When microbursts happen today, operators have few tools at their disposal to detect, diagnose and resolve them. 
Current state of the art approaches in the data plane are coarse (\eg, manually pausing, or evicting a workload), or defer the choice to end-to-end congestion control protocols, which base their decisions on local self-interested policies without knowledge of a global objective or preferences. 
If, on the other hand, an INT report collector has reports from every packet passing through a queue, it can rebuild the state of the queue at every time instance. 
First, it can detect when the queue is heavily congested and exceeds a threshold; second, it can identify which flows contributed to the event, including those taking the lion's share. 
It can check if a flow is entitled to its share, then resolve the problem by, for example, throttling flows at their source end hosts or rerouting victim flows so that they are less affected by the issue.
Doing this well requires global knowledge and coordination within a few RTTs, which is the role of the reflex plane.

A reflex plane running on a nanoPU can perform more powerful algorithms than today's programmable switches; we can introduce intelligent algorithms, policy checking, or even AI-based algorithms.

\paragraph{b) Tuning Forwarding Plane Parameters.} If we want a network to maintain high load and low latency, we often need to tune forwarding plane parameters. 
This is because most congestion control algorithms were designed with only local information: Basic TCP makes an end host-local decision, with a simple objective such as global fairness, but without the ability to implement other global policies. 
In the network, we use ECN, DCQCN, and PFC messages to make crude decisions based on local information. This is because we generally assume that fast decisions must run in fixed function hardware. 
But, if we want the network to work well, we need to update ECN marking thresholds, buffer configuration parameters, and PFC parameters frequently as workloads constantly change.
For example, it is well known that DCQCN performs better if the parameters are tuned for specific traffic workloads (message sizes, sensitivity to completion time, etc.)~\cite{dcqcn}.
A fast reflex plane can monitor workload changes over time, and derive optimal parameters for each device and end host, providing a middle ground between fast, crude, local decisions in switches, or slow traffic engineering decisions in the control plane.

\begin{table*}[]
\begin{center}
\small
\begin{tabular}{lll}
\toprule
\textbf{}                                                                                                         & \textbf{Description}                                                                                                                                                                                                                                                                                                                                                                  & \textbf{\begin{tabular}[c]{@{}l@{}}Core challenges\\ with today's approach\end{tabular}}                                                                                                                                                       \\ 
\midrule
\textbf{\begin{tabular}[c]{@{}l@{}}Microburst detection \\ and mitigation\end{tabular}}                           & \begin{tabular}[c]{@{}l@{}}Detect increase in queueing delay, identify \\ culprit flows, and take corrective action.\end{tabular}                                                                                                                                                                                                                                                     & \begin{tabular}[c]{@{}l@{}}Quickly detecting and resolving it before \\ it starts causing damages to victim flows.\end{tabular}                                                                                                                \\ \hline
\textbf{\begin{tabular}[c]{@{}l@{}}Data plane parameter\\ learning and tuning\end{tabular}}                       & \begin{tabular}[c]{@{}l@{}}Use telemetry measurements to configure \\ optimal data plane parameters for varying\\ environments.\end{tabular}                                                                                                                                                                                                                                          & \begin{tabular}[c]{@{}l@{}}Quickly adapting to varying traffic\\ patterns, applications, and requirements\end{tabular}                                                                                                                         \\ \hline
\textbf{\begin{tabular}[c]{@{}l@{}}DDoS/super-spreader/\\ port-scanning \\ detection and mitigation\end{tabular}} & \begin{tabular}[c]{@{}l@{}}Detect when many different connections are\\ opened on the same host or by the same host\\ within a short period of time. Configure \\ firewall rules to block additional connections.\end{tabular}                                                                                                                                                        & \begin{tabular}[c]{@{}l@{}}Ensuring extremely low detection\\ latency to minimize damage. \\ Detecting highly-distributed attacks\\ that can evade local detection.\end{tabular}                                                               \\ \hline
\textbf{Distributed rate limiting}                                                                                & \begin{tabular}[c]{@{}l@{}}Detect when a collection of VMs, containers, \\ applications, or tenants attempts to consume\\ more than its fair share of network resources.\\ Apply rate limiting in a distributed fashion \\ and keep adjusting the local rate caps to avoid \\ unnecessary throttling as long as the aggregate\\ consumption is less than the fair share.\end{tabular} & \begin{tabular}[c]{@{}l@{}}Detecting a violation at the aggregate\\ level. Distributing available resources \\ per desired global policy (e.g., Max-min\\ fairness) without introducing any \\ unnecessary suboptimal allocation.\end{tabular} \\ \hline
\textbf{\begin{tabular}[c]{@{}l@{}}Routing loop detection\\ and mitigation\end{tabular}}                          & \begin{tabular}[c]{@{}l@{}}Monitor the path that each packet takes \\ through the network and look for loops.\\ Update routing rules to resolve it\end{tabular}                                                                                                                                                                                                                       & \begin{tabular}[c]{@{}l@{}}Detecting transient loops timely and \\ responding to them by rerouting \\ only necessary traffic while taking into \\ account global topology and routing policy.\end{tabular}                                            \\ 
\bottomrule
\end{tabular}
\caption{Example reflex operations.}
\label{tab:reflex-examples}
\vspace{-25pt}
\end{center}
\end{table*}


\subsection{Design of a Reflex Plane }
\label{sec:reflex-design}

\begin{figure}
  \centering
  \includegraphics[width=1\linewidth]{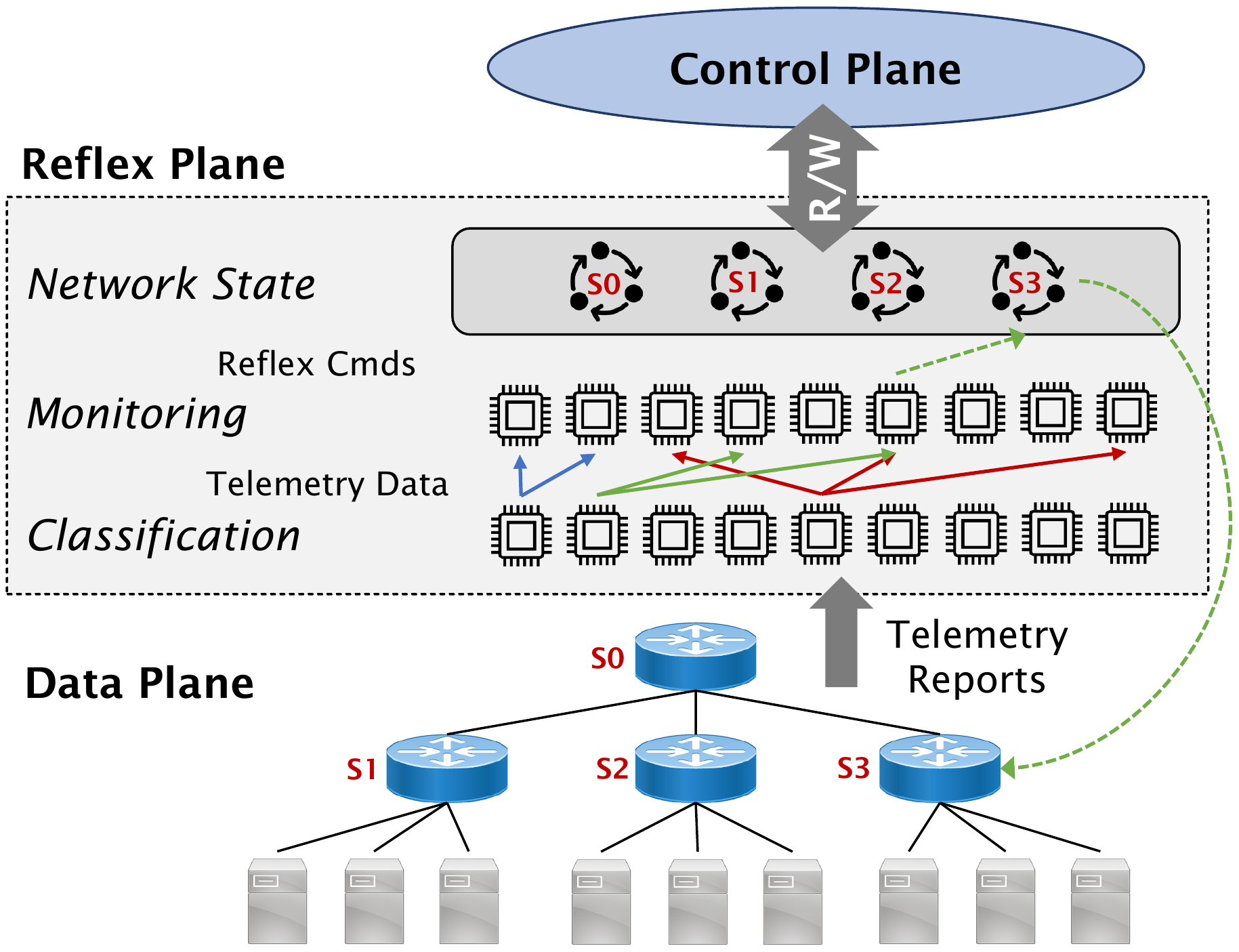}
  \vspace{-15pt}
  \caption{The design of reflex-plane architecture.}
  \label{fig:reflex-arch}
\end{figure}

Figure~\ref{fig:reflex-arch} is a high-level diagram of our reflex plane architecture.
The overall goal of the reflex plane is to enable fine-grained network monitoring with extremely low latency (real-time) closed loop control.
This section explains each component of the architecture in detail.
Note that optimizing for latency through each component is paramount if we are to minimize the overall control loop latency.

\paragraph{Telemetry Reports.} The data plane provides telemetry reports containing information about the current state of the network.
We will assume that INT~\cite{INT} reports are used, as described in Section~\ref{sec:telemetry-service}.
The INT sink (the last switch along the path) extracts the INT header and metadata from each packet, delivers the original packet to the destination host, and forwards the INT metadata to the reflex plane. 
Each switch additionally reports packet drops to the reflex plane, along with the reason for the drop and switch metadata from previous hops. Duplicate reports are coalesced or removed.

\paragraph{Classification Layer.} The job of the classification layer is to examine the telemetry reports and decide which reflex processing nodes to send the data to for further processing and analysis. 
A given report may need to be replicated and then sent to multiple processing nodes for scaling or for multiple concurrent monitoring functions each of which is responsible for monitoring particular events based on its own state. 
Arriving telemetry reports are classified based upon a number of fields, including: flow identification fields (\eg, 5 tuple), switch \& link ID, switch \& queue ID, and the reason for a packet drop.
The classification rules may include wildcards, range matches, and a priority value for each rule.
The rule indicates which report fields should be sent to reflex plane monitoring nodes for further processing. The classification rules need to be updated when monitoring applications are added and removed.
The classification layer is easily parallelizable.
It should be provisioned with enough capacity in a scaled-out fashion to classify all telemetry reports from the data plane because, as we mentioned earlier, a \SI{100}{GE} line carrying 1500B packets can generate 8.3M reports/second. 

\vspace{-2pt}
\paragraph{Monitoring Layer.} The monitors in the monitoring layer are where decisions are made, and contain the main logic of the reflex plane.
Each monitor runs on a dedicated set of nanoPU cores and processes the telemetry metadata sent to it by the classifier.
A monitor must maintain its own state; for example, flow, queue, or link state needed for its decisions. Monitor state is assumed to be ephemeral, based on recent measurements and reports, and does not need to be fault tolerant.
The number of nanoPU nodes needed for the monitoring layer depends on the number of monitors and the number of incoming telemetry reports. 
The system needs to scale dynamically as needed, as demand grows. 
Each monitoring application is self-contained, keeping track of the state it needs, allowing it to scale out by provisioning additional nanoPU cores on demand.

The amount of processing a monitoring node can perform depends on whether it maintains state. 
Using our \SI{100}{GE} link as an example, carrying 1500B and 8.3M reports/second, if a single monitoring node needs to process every report, then is must complete within \SI{120}{ns}, or about 400 instructions on a \SI{3.2}{GHz} nanoPU. 
This, we have found, is often sufficient. However, if the monitor needs more cycles, it can further pipeline the function across several nanoPU cores, so long as state is confined to one core, or carried with the data. 
If the monitor is stateless (\eg, if its job is to simply compare a telemetry report against a set of fixed thresholds) then the classification layer can be told to load-balance reports over multiple identical instances to scale out performance.

When a monitor makes a decision, it produces reflex commands authorized by the control plane; for example, to update switch state to re-route a flow or tune an ECN marking threshold. 
Reflex commands are sent directly to the subsequent layer in the reflex plane, the network state layer.


\paragraph{Network State Layer.} The network state layer maintains the state associated with each network element in a reliable, replicated store. 
We assume here that it uses Raft~\cite{raft}, but it could use any consensus protocol that enables replicated state machines. 
Modern distributed SDN controllers already maintain replicated state internally; we propose refactoring the design so as to offload some of this state into the reflex plane where it can be updated more quickly.

We need to be clear about which state needs to be fault tolerant, and which does not. 
Generally, each monitor maintains ephemeral internal state (\eg, counters) that do not need to be redundant. 
If a monitoring node fails, it starts over. 
However, state that configures the monitor, \eg, a threshold value, or configuration that the monitor updates, such as the forwarding rules in a switch, is part of the global network state owned by the network state layer of the reflex plane. 
This state is treated with the same importance as the control plane's internal replicated state and must survive failures of individual components in the reflex plane. 
Clearly identifying which state needs to be replicated, we can keep monitoring applications simple and scalable. 
But this must not be at the expense of overall global state reliability.

When a monitor sends a reflex command that updates global state, it sends it to the nanoPU-based Raft cluster responsible for maintaining the state associated with the target network elements.
The cluster applies the reflex command to its state and subsequently forwards the command to the physical switch.
This ensures that the control plane will always have a consistent view of the relevant state in the forwarding plane.


\paragraph{Control-Plane Interface.} The control plane is able to read the state of each network element in the reflex plane.
SDN controllers use this state to maintain a global view of the network and check properties, such as reachability between hosts. 
Monitors in the reflex plane may query some state within the control plane, such as the network topology.
Therefore, we recommend that this global state is exposed by a very low-latency RPC service, allowing the reflex plane to react quickly.
The control plane can also write the state of each network element by sending control commands to the reflex plane. 
For example, if a control program implements BGP, it will write new forwarding entries into the switch tables. These are stored redundantly by the reflex plane for fast, reliable access.

\subsection{Discussion}
The performance of the reflex plane is dictated by the performance of each layer in the usual way. 
The latency will be additive across the layers; the throughput will be determined by the bottleneck. 
Our evaluations in Section~\ref{sec:eval} suggest that it is feasible to build a reflex plane with an overall reflex time below \SI{10}{$\mu$}s when using nanoPUs. 
We believe it is much faster than any existing network control planes.

\section{Related Work}

\paragraph{Packet Classification.}
High performance packet classification is a hot topic of networking research.
The highest performing packet classification systems use dedicated hardware which is specifically designed for this task, such as the PISA~\cite{RMT} switch architecture.
These devices can classify packets at more than 10 Tb/s, more than 15 billion packets per second.
Software packet classification techniques~\cite{nuevomatch, cutsplit, tuplemerge} are commonly used in virtual network functions, such as forwarders or ACL firewalls.
These approaches classify packets at a few million packets per second.
Rather than proposing a new technique for packet classification, we explore how it can be accelerating using the recently proposed nanoPU.

\paragraph{Network Monitoring Systems.}
The advent of In-band Network Telemetry (INT) and programmable data planes has sparked a renewed interest in how we perform network monitoring.
INT Collector~\cite{intcollector} and Barefoot/Intel Deep Insight~\cite{deep-insight} are two existing systems to process INT reports and store the results in a time series database so that network operators can run queries to diagnose performance issues.
Sonata~\cite{sonata}, Marple~\cite{marple}, UnivMon~\cite{univmon}, and Speedlight~\cite{speedlight} are proposals that harness data plane programmability to enable more scalable monitoring systems, capable of handling the firehose of telemetry reports.
However, unlike our proposed reflex plane, none of these systems have the explicit goal of enabling real time detection and mitigation of observed issues.

\paragraph{State Machine Replication.}
NetChain~\cite{netchain} and NetPaxos~\cite{netpaxos} are two systems that leverage data-plane programmability to accelerate state machine replication logic.
NetChain seeks to accelerate chain replication and NetPaxos accelerates Paxos~\cite{paxos}.
We explore how the nanoPU can be used to accelerate the Raft consensus protocol~\cite{raft}.

\paragraph{Low Latency Network Control.}
Mantis~\cite{mantis}, whose goal is to enable low latency control plane response times, was one of our original inspirations for the reflex plane.
The Mantis control plane runs on the switches' local CPUs and continuously polls the switch state. It runs ``reaction functions'', and updates the switch state.
This approach is quite limited, however, because switches do not have access to global information about the network and hence the reaction functions must operate with limited knowledge.
Furthermore, we believe it would be challenging to integrate Mantis into an existing SDN because there is no mechanism for Mantis to communicate relevant state updates to the central controller.
We propose the reflex plane in an attempt to solve these issues.


\paragraph{Redesigning the network abstraction layers.}
We are not the first to suggest revisiting the layers of abstraction used to define network systems.
4D~\cite{4D} is a precursor to the modern notion of software defined networking.
It advocates for refactoring the decision logic that is baked into the network elements into a separate data plane and decision plane, connected by discovery and dissemination planes.
This idea was embraced by SDN where the control plane was moved out of network elements and into a logically centralized controller.
The reflex plane shares some common goals with the knowledge plane~\cite{knowledge-plane}, which was described in 2003.
In particular, both proposals advocate for the need to make network control more automated and intelligent.

\section{Conclusion}

In the past, we had at our disposal two classes of processing elements for networks: The control plane, invariably running in software on one or more CPUs; and the forwarding plane, typically running on a fixed-function ASIC. 
It was natural to think of network functions as ``complex and sophisticated, but slow'' (running in software) or ``dumb, simple and fast'' (running in switch hardware). 
With the advent of programmable switches, the stark line has become blurred, allowing us to run more sophisticated, stateful control operations at line rate in the forwarding plane. 
Over time, as more programmable switches and NICs become available, it will be tempting to place more control functions into the forwarding plane. 
When doing so, we will almost certainly continue to wrestle with the desire for fast control loops, pushing us towards the forwarding plane, versus the richness of the control operations and the redundancy of the state, pushing us towards software in the control plane.

We therefore think it is important to keep evaluating new accelerators and domain-specific processors, to see if they offer an intermediate solution. 
With Moore's Law slowing down, more hardware accelerators will appear. 
The nanoPU itself is one example of this trend. 
With its high throughput, low and predictable latency communication model, coupled with a familiar CPU core instruction set, the nanoPU provides an interesting opportunity to place very fast reflex decisions in the network, based on very fine grained telemetry data from the forwarding plane, while allowing monitor functions to scale and easily evolve and improve over time.

\end{sloppypar}

\label{lastpage}

\if\showacks1
    \begin{acks}
To Robert, for the bagels and explaining CMYK and color spaces.
\end{acks}

\fi

\bibliographystyle{acmbib}
\bibliography{bibs/paper}


\begin{thebibliography}{57}


\ifx \showCODEN    \undefined \def \showCODEN     #1{\unskip}     \fi
\ifx \showDOI      \undefined \def \showDOI       #1{#1}\fi
\ifx \showISBNx    \undefined \def \showISBNx     #1{\unskip}     \fi
\ifx \showISBNxiii \undefined \def \showISBNxiii  #1{\unskip}     \fi
\ifx \showISSN     \undefined \def \showISSN      #1{\unskip}     \fi
\ifx \showLCCN     \undefined \def \showLCCN      #1{\unskip}     \fi
\ifx \shownote     \undefined \def \shownote      #1{#1}          \fi
\ifx \showarticletitle \undefined \def \showarticletitle #1{#1}   \fi
\ifx \showURL      \undefined \def \showURL       {\relax}        \fi
\providecommand\bibfield[2]{#2}
\providecommand\bibinfo[2]{#2}
\providecommand\natexlab[1]{#1}
\providecommand\showeprint[2][]{arXiv:#2}

\bibitem[\protect\citeauthoryear{Arslan, Ibanez, Mallery, Kim, and
  McKeown}{Arslan et~al\mbox{.}}{2021}]%
        {nanotransport}
\bibfield{author}{\bibinfo{person}{Serhat Arslan}, \bibinfo{person}{Stephen
  Ibanez}, \bibinfo{person}{Alex Mallery}, \bibinfo{person}{Changhoon Kim},
  {and} \bibinfo{person}{Nick McKeown}.} \bibinfo{year}{2021}\natexlab{}.
\newblock \showarticletitle{NanoTransport: A Low-Latency, Programmable
  Transport Layer for NICs}. In \bibinfo{booktitle}{\emph{Proceedings of the
  ACM SIGCOMM Symposium on SDN Research (SOSR)}} (Virtual Event, USA)
  \emph{(\bibinfo{series}{SOSR '21})}. \bibinfo{publisher}{Association for
  Computing Machinery}, \bibinfo{address}{New York, NY, USA},
  \bibinfo{pages}{13–26}.
\newblock
\showISBNx{9781450390842}
\urldef\tempurl%
\url{https://doi.org/10.1145/3482898.3483365}
\showDOI{\tempurl}


\bibitem[\protect\citeauthoryear{Asanovic, Avizienis, Bachrach, Beamer,
  Biancolin, Celio, Cook, Dabbelt, Hauser, Izraelevitz, et~al\mbox{.}}{Asanovic
  et~al\mbox{.}}{2016}]%
        {rocket-chip}
\bibfield{author}{\bibinfo{person}{Krste Asanovic}, \bibinfo{person}{Rimas
  Avizienis}, \bibinfo{person}{Jonathan Bachrach}, \bibinfo{person}{Scott
  Beamer}, \bibinfo{person}{David Biancolin}, \bibinfo{person}{Christopher
  Celio}, \bibinfo{person}{Henry Cook}, \bibinfo{person}{Daniel Dabbelt},
  \bibinfo{person}{John Hauser}, \bibinfo{person}{Adam Izraelevitz},
  {et~al\mbox{.}}} \bibinfo{year}{2016}\natexlab{}.
\newblock \showarticletitle{The rocket chip generator}.
\newblock \bibinfo{journal}{\emph{EECS Department, University of California,
  Berkeley, Tech. Rep. UCB/EECS-2016-17}} (\bibinfo{year}{2016}).
\newblock


\bibitem[\protect\citeauthoryear{Bosshart, Gibb, Kim, Varghese, McKeown,
  Izzard, Mujica, and Horowitz}{Bosshart et~al\mbox{.}}{2013}]%
        {RMT}
\bibfield{author}{\bibinfo{person}{Pat Bosshart}, \bibinfo{person}{Glen Gibb},
  \bibinfo{person}{Hun-Seok Kim}, \bibinfo{person}{George Varghese},
  \bibinfo{person}{Nick McKeown}, \bibinfo{person}{Martin Izzard},
  \bibinfo{person}{Fernando Mujica}, {and} \bibinfo{person}{Mark Horowitz}.}
  \bibinfo{year}{2013}\natexlab{}.
\newblock \showarticletitle{Forwarding metamorphosis: Fast programmable
  match-action processing in hardware for SDN}.
\newblock \bibinfo{journal}{\emph{ACM SIGCOMM Computer Communication Review}}
  \bibinfo{volume}{43}, \bibinfo{number}{4} (\bibinfo{year}{2013}),
  \bibinfo{pages}{99--110}.
\newblock


\bibitem[\protect\citeauthoryear{Burrows}{Burrows}{2006}]%
        {chubby}
\bibfield{author}{\bibinfo{person}{Mike Burrows}.}
  \bibinfo{year}{2006}\natexlab{}.
\newblock \showarticletitle{The Chubby lock service for loosely-coupled
  distributed systems}. In \bibinfo{booktitle}{\emph{Proceedings of the 7th
  symposium on Operating systems design and implementation}}.
  \bibinfo{pages}{335--350}.
\newblock


\bibitem[\protect\citeauthoryear{Chen, Feibish, Koral, Rexford, Rottenstreich,
  Monetti, and Wang}{Chen et~al\mbox{.}}{2019}]%
        {conquest}
\bibfield{author}{\bibinfo{person}{Xiaoqi Chen}, \bibinfo{person}{Shir~Landau
  Feibish}, \bibinfo{person}{Yaron Koral}, \bibinfo{person}{Jennifer Rexford},
  \bibinfo{person}{Ori Rottenstreich}, \bibinfo{person}{Steven~A Monetti},
  {and} \bibinfo{person}{Tzuu-Yi Wang}.} \bibinfo{year}{2019}\natexlab{}.
\newblock \showarticletitle{Fine-grained queue measurement in the data plane}.
  In \bibinfo{booktitle}{\emph{Proceedings of the 15th International Conference
  on Emerging Networking Experiments And Technologies}}.
  \bibinfo{pages}{15--29}.
\newblock


\bibitem[\protect\citeauthoryear{Chen, Wei, Shi, Chen, and Chen}{Chen
  et~al\mbox{.}}{2016}]%
        {drtmr}
\bibfield{author}{\bibinfo{person}{Yanzhe Chen}, \bibinfo{person}{Xingda Wei},
  \bibinfo{person}{Jiaxin Shi}, \bibinfo{person}{Rong Chen}, {and}
  \bibinfo{person}{Haibo Chen}.} \bibinfo{year}{2016}\natexlab{}.
\newblock \showarticletitle{Fast and general distributed transactions using
  RDMA and HTM}. In \bibinfo{booktitle}{\emph{Proceedings of the Eleventh
  European Conference on Computer Systems}}. \bibinfo{pages}{1--17}.
\newblock


\bibitem[\protect\citeauthoryear{Clark, Partridge, Ramming, and
  Wroclawski}{Clark et~al\mbox{.}}{2003}]%
        {knowledge-plane}
\bibfield{author}{\bibinfo{person}{David~D Clark}, \bibinfo{person}{Craig
  Partridge}, \bibinfo{person}{J~Christopher Ramming}, {and}
  \bibinfo{person}{John~T Wroclawski}.} \bibinfo{year}{2003}\natexlab{}.
\newblock \showarticletitle{A knowledge plane for the internet}. In
  \bibinfo{booktitle}{\emph{Proceedings of the 2003 conference on Applications,
  technologies, architectures, and protocols for computer communications}}.
  \bibinfo{pages}{3--10}.
\newblock


\bibitem[\protect\citeauthoryear{Daly, Bruschi, Linguaglossa, Pontarelli,
  Rossi, Tollet, Torng, and Yourtchenko}{Daly et~al\mbox{.}}{2019}]%
        {tuplemerge}
\bibfield{author}{\bibinfo{person}{James Daly}, \bibinfo{person}{Valerio
  Bruschi}, \bibinfo{person}{Leonardo Linguaglossa}, \bibinfo{person}{Salvatore
  Pontarelli}, \bibinfo{person}{Dario Rossi}, \bibinfo{person}{Jerome Tollet},
  \bibinfo{person}{Eric Torng}, {and} \bibinfo{person}{Andrew Yourtchenko}.}
  \bibinfo{year}{2019}\natexlab{}.
\newblock \showarticletitle{Tuplemerge: Fast software packet processing for
  online packet classification}.
\newblock \bibinfo{journal}{\emph{IEEE/ACM transactions on networking}}
  \bibinfo{volume}{27}, \bibinfo{number}{4} (\bibinfo{year}{2019}),
  \bibinfo{pages}{1417--1431}.
\newblock


\bibitem[\protect\citeauthoryear{Dang, Sciascia, Canini, Pedone, and
  Soul{\'e}}{Dang et~al\mbox{.}}{2015}]%
        {netpaxos}
\bibfield{author}{\bibinfo{person}{Huynh~Tu Dang}, \bibinfo{person}{Daniele
  Sciascia}, \bibinfo{person}{Marco Canini}, \bibinfo{person}{Fernando Pedone},
  {and} \bibinfo{person}{Robert Soul{\'e}}.} \bibinfo{year}{2015}\natexlab{}.
\newblock \showarticletitle{Netpaxos: Consensus at network speed}. In
  \bibinfo{booktitle}{\emph{Proceedings of the 1st ACM SIGCOMM Symposium on
  Software Defined Networking Research}}. \bibinfo{pages}{1--7}.
\newblock


\bibitem[\protect\citeauthoryear{Dean and Barroso}{Dean and Barroso}{2013}]%
        {tail-at-scale}
\bibfield{author}{\bibinfo{person}{Jeffrey Dean} {and}
  \bibinfo{person}{Luiz~Andr{\'e} Barroso}.} \bibinfo{year}{2013}\natexlab{}.
\newblock \showarticletitle{The tail at scale}.
\newblock \bibinfo{journal}{\emph{Commun. ACM}} \bibinfo{volume}{56},
  \bibinfo{number}{2} (\bibinfo{year}{2013}), \bibinfo{pages}{74--80}.
\newblock


\bibitem[\protect\citeauthoryear{Deep-Insight}{Deep-Insight}{[n.d.]}]%
        {deep-insight}
Deep-Insight \bibinfo{year}{[n.d.]}\natexlab{}.
\newblock \bibinfo{title}{Barefoot/Intel Deep Insight}.
\newblock
  \bibinfo{howpublished}{\url{https://www.barefootnetworks.com/products/brief-deep-insight/}}.
\newblock
\newblock
\shownote{Accessed on 09/15/2020.}


\bibitem[\protect\citeauthoryear{Greenberg, Hjalmtysson, Maltz, Myers, Rexford,
  Xie, Yan, Zhan, and Zhang}{Greenberg et~al\mbox{.}}{2005}]%
        {4D}
\bibfield{author}{\bibinfo{person}{Albert Greenberg}, \bibinfo{person}{Gisli
  Hjalmtysson}, \bibinfo{person}{David~A Maltz}, \bibinfo{person}{Andy Myers},
  \bibinfo{person}{Jennifer Rexford}, \bibinfo{person}{Geoffrey Xie},
  \bibinfo{person}{Hong Yan}, \bibinfo{person}{Jibin Zhan}, {and}
  \bibinfo{person}{Hui Zhang}.} \bibinfo{year}{2005}\natexlab{}.
\newblock \showarticletitle{A clean slate 4D approach to network control and
  management}.
\newblock \bibinfo{journal}{\emph{ACM SIGCOMM Computer Communication Review}}
  \bibinfo{volume}{35}, \bibinfo{number}{5} (\bibinfo{year}{2005}),
  \bibinfo{pages}{41--54}.
\newblock


\bibitem[\protect\citeauthoryear{Gupta, Harrison, Canini, Feamster, Rexford,
  and Willinger}{Gupta et~al\mbox{.}}{2018}]%
        {sonata}
\bibfield{author}{\bibinfo{person}{Arpit Gupta}, \bibinfo{person}{Rob
  Harrison}, \bibinfo{person}{Marco Canini}, \bibinfo{person}{Nick Feamster},
  \bibinfo{person}{Jennifer Rexford}, {and} \bibinfo{person}{Walter
  Willinger}.} \bibinfo{year}{2018}\natexlab{}.
\newblock \showarticletitle{Sonata: Query-driven streaming network telemetry}.
  In \bibinfo{booktitle}{\emph{Proceedings of the 2018 Conference of the ACM
  Special Interest Group on Data Communication}}. \bibinfo{pages}{357--371}.
\newblock


\bibitem[\protect\citeauthoryear{Gupta and McKeown}{Gupta and McKeown}{2001}]%
        {packetClassification}
\bibfield{author}{\bibinfo{person}{P. Gupta} {and} \bibinfo{person}{N.
  McKeown}.} \bibinfo{year}{2001}\natexlab{}.
\newblock \showarticletitle{Algorithms for Packet Classification}.
\newblock \bibinfo{journal}{\emph{Netwrk. Mag. of Global Internetwkg.}}
  \bibinfo{volume}{15}, \bibinfo{number}{2} (\bibinfo{date}{March}
  \bibinfo{year}{2001}), \bibinfo{pages}{24–32}.
\newblock
\showISSN{0890-8044}
\urldef\tempurl%
\url{https://doi.org/10.1109/65.912717}
\showDOI{\tempurl}


\bibitem[\protect\citeauthoryear{Handley, Raiciu, Agache, Voinescu, Moore,
  Antichi, and W{\'o}jcik}{Handley et~al\mbox{.}}{2017}]%
        {ndp}
\bibfield{author}{\bibinfo{person}{Mark Handley}, \bibinfo{person}{Costin
  Raiciu}, \bibinfo{person}{Alexandru Agache}, \bibinfo{person}{Andrei
  Voinescu}, \bibinfo{person}{Andrew~W Moore}, \bibinfo{person}{Gianni
  Antichi}, {and} \bibinfo{person}{Marcin W{\'o}jcik}.}
  \bibinfo{year}{2017}\natexlab{}.
\newblock \showarticletitle{Re-architecting datacenter networks and stacks for
  low latency and high performance}. In \bibinfo{booktitle}{\emph{Proceedings
  of the Conference of the ACM Special Interest Group on Data Communication}}.
  \bibinfo{pages}{29--42}.
\newblock


\bibitem[\protect\citeauthoryear{Hsu, Beckett, Chen, Rexford, and Walker}{Hsu
  et~al\mbox{.}}{2020}]%
        {contra}
\bibfield{author}{\bibinfo{person}{Kuo-Feng Hsu}, \bibinfo{person}{Ryan
  Beckett}, \bibinfo{person}{Ang Chen}, \bibinfo{person}{Jennifer Rexford},
  {and} \bibinfo{person}{David Walker}.} \bibinfo{year}{2020}\natexlab{}.
\newblock \showarticletitle{Contra: A programmable system for performance-aware
  routing}. In \bibinfo{booktitle}{\emph{17th $\{$USENIX$\}$ Symposium on
  Networked Systems Design and Implementation ($\{$NSDI$\}$ 20)}}.
  \bibinfo{pages}{701--721}.
\newblock


\bibitem[\protect\citeauthoryear{Hunt, Konar, Junqueira, and Reed}{Hunt
  et~al\mbox{.}}{2010}]%
        {zookeeper}
\bibfield{author}{\bibinfo{person}{Patrick Hunt}, \bibinfo{person}{Mahadev
  Konar}, \bibinfo{person}{Flavio~Paiva Junqueira}, {and}
  \bibinfo{person}{Benjamin Reed}.} \bibinfo{year}{2010}\natexlab{}.
\newblock \showarticletitle{ZooKeeper: Wait-free Coordination for
  Internet-scale Systems.}. In \bibinfo{booktitle}{\emph{USENIX annual
  technical conference}}, Vol.~\bibinfo{volume}{8}.
\newblock


\bibitem[\protect\citeauthoryear{Ibanez, Mallery, Arslan, Jepsen, Shahbaz,
  McKeown, and Kim}{Ibanez et~al\mbox{.}}{2021}]%
        {nanopu}
\bibfield{author}{\bibinfo{person}{Stephen Ibanez}, \bibinfo{person}{Alex
  Mallery}, \bibinfo{person}{Serhat Arslan}, \bibinfo{person}{Theo Jepsen},
  \bibinfo{person}{Muhammad Shahbaz}, \bibinfo{person}{Nick McKeown}, {and}
  \bibinfo{person}{Changhoon Kim}.} \bibinfo{year}{2021}\natexlab{}.
\newblock \showarticletitle{The nanoPU: A Nanosecond Network Stack for
  Datacenters}. In \bibinfo{booktitle}{\emph{15th {USENIX} Symposium on
  Operating Systems Design and Implementation ({OSDI} 21)}}.
  \bibinfo{publisher}{{USENIX} Association}, \bibinfo{address}{Boston, MA}.
\newblock
\urldef\tempurl%
\url{https://www.usenix.org/conference/osdi21/presentation/ibanez}
\showURL{%
\tempurl}


\bibitem[\protect\citeauthoryear{Ibanez, Shahbaz, and McKeown}{Ibanez
  et~al\mbox{.}}{2019}]%
        {lnic}
\bibfield{author}{\bibinfo{person}{Stephen Ibanez}, \bibinfo{person}{Muhammad
  Shahbaz}, {and} \bibinfo{person}{Nick McKeown}.}
  \bibinfo{year}{2019}\natexlab{}.
\newblock \showarticletitle{The Case for a Network Fast Path to the CPU}. In
  \bibinfo{booktitle}{\emph{Proceedings of the 18th ACM Workshop on Hot Topics
  in Networks}}. \bibinfo{pages}{52--59}.
\newblock


\bibitem[\protect\citeauthoryear{Jin, Li, Zhang, Foster, Lee, Soul{\'e}, Kim,
  and Stoica}{Jin et~al\mbox{.}}{2018}]%
        {netchain}
\bibfield{author}{\bibinfo{person}{Xin Jin}, \bibinfo{person}{Xiaozhou Li},
  \bibinfo{person}{Haoyu Zhang}, \bibinfo{person}{Nate Foster},
  \bibinfo{person}{Jeongkeun Lee}, \bibinfo{person}{Robert Soul{\'e}},
  \bibinfo{person}{Changhoon Kim}, {and} \bibinfo{person}{Ion Stoica}.}
  \bibinfo{year}{2018}\natexlab{}.
\newblock \showarticletitle{Netchain: Scale-free sub-rtt coordination}. In
  \bibinfo{booktitle}{\emph{15th $\{$USENIX$\}$ Symposium on Networked Systems
  Design and Implementation ($\{$NSDI$\}$ 18)}}. \bibinfo{pages}{35--49}.
\newblock


\bibitem[\protect\citeauthoryear{Jin, Li, Zhang, Soul{\'e}, Lee, Foster, Kim,
  and Stoica}{Jin et~al\mbox{.}}{2017}]%
        {netcache}
\bibfield{author}{\bibinfo{person}{Xin Jin}, \bibinfo{person}{Xiaozhou Li},
  \bibinfo{person}{Haoyu Zhang}, \bibinfo{person}{Robert Soul{\'e}},
  \bibinfo{person}{Jeongkeun Lee}, \bibinfo{person}{Nate Foster},
  \bibinfo{person}{Changhoon Kim}, {and} \bibinfo{person}{Ion Stoica}.}
  \bibinfo{year}{2017}\natexlab{}.
\newblock \showarticletitle{Netcache: Balancing key-value stores with fast
  in-network caching}. In \bibinfo{booktitle}{\emph{Proceedings of the 26th
  Symposium on Operating Systems Principles}}. \bibinfo{pages}{121--136}.
\newblock


\bibitem[\protect\citeauthoryear{Kaffes, Chong, Humphries, Belay, Mazi{\`e}res,
  and Kozyrakis}{Kaffes et~al\mbox{.}}{2019}]%
        {shinjuku}
\bibfield{author}{\bibinfo{person}{Kostis Kaffes}, \bibinfo{person}{Timothy
  Chong}, \bibinfo{person}{Jack~Tigar Humphries}, \bibinfo{person}{Adam Belay},
  \bibinfo{person}{David Mazi{\`e}res}, {and} \bibinfo{person}{Christos
  Kozyrakis}.} \bibinfo{year}{2019}\natexlab{}.
\newblock \showarticletitle{Shinjuku: Preemptive scheduling for
  $\mu$second-scale tail latency}. In \bibinfo{booktitle}{\emph{16th
  $\{$USENIX$\}$ Symposium on Networked Systems Design and Implementation
  ($\{$NSDI$\}$ 19)}}. \bibinfo{pages}{345--360}.
\newblock


\bibitem[\protect\citeauthoryear{Kalia, Kaminsky, and Andersen}{Kalia
  et~al\mbox{.}}{2019}]%
        {eRPC}
\bibfield{author}{\bibinfo{person}{Anuj Kalia}, \bibinfo{person}{Michael
  Kaminsky}, {and} \bibinfo{person}{David Andersen}.}
  \bibinfo{year}{2019}\natexlab{}.
\newblock \showarticletitle{Datacenter RPCs can be general and fast}. In
  \bibinfo{booktitle}{\emph{16th $\{$USENIX$\}$ Symposium on Networked Systems
  Design and Implementation ($\{$NSDI$\}$ 19)}}. \bibinfo{pages}{1--16}.
\newblock


\bibitem[\protect\citeauthoryear{Kalia, Kaminsky, and Andersen}{Kalia
  et~al\mbox{.}}{2014}]%
        {herd}
\bibfield{author}{\bibinfo{person}{Anuj Kalia}, \bibinfo{person}{Michael
  Kaminsky}, {and} \bibinfo{person}{David~G Andersen}.}
  \bibinfo{year}{2014}\natexlab{}.
\newblock \showarticletitle{Using RDMA efficiently for key-value services}. In
  \bibinfo{booktitle}{\emph{Proceedings of the 2014 ACM conference on
  SIGCOMM}}. \bibinfo{pages}{295--306}.
\newblock


\bibitem[\protect\citeauthoryear{Kalia, Kaminsky, and Andersen}{Kalia
  et~al\mbox{.}}{2016}]%
        {fasst}
\bibfield{author}{\bibinfo{person}{Anuj Kalia}, \bibinfo{person}{Michael
  Kaminsky}, {and} \bibinfo{person}{David~G Andersen}.}
  \bibinfo{year}{2016}\natexlab{}.
\newblock \showarticletitle{FaSST: Fast, Scalable and Simple Distributed
  Transactions with Two-Sided ($\{$RDMA$\}$) Datagram RPCs}. In
  \bibinfo{booktitle}{\emph{12th $\{$USENIX$\}$ Symposium on Operating Systems
  Design and Implementation ($\{$OSDI$\}$ 16)}}. \bibinfo{pages}{185--201}.
\newblock


\bibitem[\protect\citeauthoryear{Karandikar, Mao, Kim, Biancolin, Amid, Lee,
  Pemberton, Amaro, Schmidt, Chopra, et~al\mbox{.}}{Karandikar
  et~al\mbox{.}}{2018}]%
        {firesim}
\bibfield{author}{\bibinfo{person}{Sagar Karandikar}, \bibinfo{person}{Howard
  Mao}, \bibinfo{person}{Donggyu Kim}, \bibinfo{person}{David Biancolin},
  \bibinfo{person}{Alon Amid}, \bibinfo{person}{Dayeol Lee},
  \bibinfo{person}{Nathan Pemberton}, \bibinfo{person}{Emmanuel Amaro},
  \bibinfo{person}{Colin Schmidt}, \bibinfo{person}{Aditya Chopra},
  {et~al\mbox{.}}} \bibinfo{year}{2018}\natexlab{}.
\newblock \showarticletitle{FireSim: FPGA-accelerated cycle-exact scale-out
  system simulation in the public cloud}. In \bibinfo{booktitle}{\emph{2018
  ACM/IEEE 45th Annual International Symposium on Computer Architecture
  (ISCA)}}. IEEE, \bibinfo{pages}{29--42}.
\newblock


\bibitem[\protect\citeauthoryear{Katta, Hira, Kim, Sivaraman, and
  Rexford}{Katta et~al\mbox{.}}{2016}]%
        {hula}
\bibfield{author}{\bibinfo{person}{Naga Katta}, \bibinfo{person}{Mukesh Hira},
  \bibinfo{person}{Changhoon Kim}, \bibinfo{person}{Anirudh Sivaraman}, {and}
  \bibinfo{person}{Jennifer Rexford}.} \bibinfo{year}{2016}\natexlab{}.
\newblock \showarticletitle{Hula: Scalable load balancing using programmable
  data planes}. In \bibinfo{booktitle}{\emph{Proceedings of the Symposium on
  SDN Research}}. \bibinfo{pages}{1--12}.
\newblock


\bibitem[\protect\citeauthoryear{Kazemian, Varghese, and McKeown}{Kazemian
  et~al\mbox{.}}{2012}]%
        {HSA}
\bibfield{author}{\bibinfo{person}{Peyman Kazemian}, \bibinfo{person}{George
  Varghese}, {and} \bibinfo{person}{Nick McKeown}.}
  \bibinfo{year}{2012}\natexlab{}.
\newblock \showarticletitle{Header space analysis: Static checking for
  networks}. In \bibinfo{booktitle}{\emph{Presented as part of the 9th
  $\{$USENIX$\}$ Symposium on Networked Systems Design and Implementation
  ($\{$NSDI$\}$ 12)}}. \bibinfo{pages}{113--126}.
\newblock


\bibitem[\protect\citeauthoryear{Kim, Sivaraman, Katta, Bas, Dixit, and
  Wobker}{Kim et~al\mbox{.}}{2015}]%
        {INT}
\bibfield{author}{\bibinfo{person}{Changhoon Kim}, \bibinfo{person}{Anirudh
  Sivaraman}, \bibinfo{person}{Naga Katta}, \bibinfo{person}{Antonin Bas},
  \bibinfo{person}{Advait Dixit}, {and} \bibinfo{person}{Lawrence~J Wobker}.}
  \bibinfo{year}{2015}\natexlab{}.
\newblock \showarticletitle{In-band network telemetry via programmable
  dataplanes}. In \bibinfo{booktitle}{\emph{ACM SIGCOMM}}.
\newblock


\bibitem[\protect\citeauthoryear{Kogias, Prekas, Ghosn, Fietz, and
  Bugnion}{Kogias et~al\mbox{.}}{2019}]%
        {r2p2}
\bibfield{author}{\bibinfo{person}{Marios Kogias}, \bibinfo{person}{George
  Prekas}, \bibinfo{person}{Adrien Ghosn}, \bibinfo{person}{Jonas Fietz}, {and}
  \bibinfo{person}{Edouard Bugnion}.} \bibinfo{year}{2019}\natexlab{}.
\newblock \showarticletitle{R2P2: Making RPCs first-class datacenter citizens}.
  In \bibinfo{booktitle}{\emph{2019 $\{$USENIX$\}$ Annual Technical Conference
  ($\{$USENIX$\}$$\{$ATC$\}$ 19)}}. \bibinfo{pages}{863--880}.
\newblock


\bibitem[\protect\citeauthoryear{Koponen, Casado, Gude, Stribling, Poutievski,
  Zhu, Ramanathan, Iwata, Inoue, Hama, et~al\mbox{.}}{Koponen
  et~al\mbox{.}}{2010}]%
        {onix}
\bibfield{author}{\bibinfo{person}{Teemu Koponen}, \bibinfo{person}{Martin
  Casado}, \bibinfo{person}{Natasha Gude}, \bibinfo{person}{Jeremy Stribling},
  \bibinfo{person}{Leon Poutievski}, \bibinfo{person}{Min Zhu},
  \bibinfo{person}{Rajiv Ramanathan}, \bibinfo{person}{Yuichiro Iwata},
  \bibinfo{person}{Hiroaki Inoue}, \bibinfo{person}{Takayuki Hama},
  {et~al\mbox{.}}} \bibinfo{year}{2010}\natexlab{}.
\newblock \showarticletitle{Onix: A distributed control platform for
  large-scale production networks.}. In \bibinfo{booktitle}{\emph{OSDI}},
  Vol.~\bibinfo{volume}{10}. \bibinfo{pages}{1--6}.
\newblock


\bibitem[\protect\citeauthoryear{Lamport et~al\mbox{.}}{Lamport
  et~al\mbox{.}}{2001}]%
        {paxos}
\bibfield{author}{\bibinfo{person}{Leslie Lamport} {et~al\mbox{.}}}
  \bibinfo{year}{2001}\natexlab{}.
\newblock \showarticletitle{Paxos made simple}.
\newblock \bibinfo{journal}{\emph{ACM Sigact News}} \bibinfo{volume}{32},
  \bibinfo{number}{4} (\bibinfo{year}{2001}), \bibinfo{pages}{18--25}.
\newblock


\bibitem[\protect\citeauthoryear{Li, Li, Li, and Xie}{Li et~al\mbox{.}}{2018}]%
        {cutsplit}
\bibfield{author}{\bibinfo{person}{Wenjun Li}, \bibinfo{person}{Xianfeng Li},
  \bibinfo{person}{Hui Li}, {and} \bibinfo{person}{Gaogang Xie}.}
  \bibinfo{year}{2018}\natexlab{}.
\newblock \showarticletitle{Cutsplit: A decision-tree combining cutting and
  splitting for scalable packet classification}. In
  \bibinfo{booktitle}{\emph{IEEE INFOCOM 2018-IEEE Conference on Computer
  Communications}}. IEEE, \bibinfo{pages}{2645--2653}.
\newblock


\bibitem[\protect\citeauthoryear{Li, Miao, Liu, Zhuang, Feng, Tang, Cao, Zhang,
  Kelly, Alizadeh, et~al\mbox{.}}{Li et~al\mbox{.}}{2019}]%
        {hpcc}
\bibfield{author}{\bibinfo{person}{Yuliang Li}, \bibinfo{person}{Rui Miao},
  \bibinfo{person}{Hongqiang~Harry Liu}, \bibinfo{person}{Yan Zhuang},
  \bibinfo{person}{Fei Feng}, \bibinfo{person}{Lingbo Tang},
  \bibinfo{person}{Zheng Cao}, \bibinfo{person}{Ming Zhang},
  \bibinfo{person}{Frank Kelly}, \bibinfo{person}{Mohammad Alizadeh},
  {et~al\mbox{.}}} \bibinfo{year}{2019}\natexlab{}.
\newblock \showarticletitle{HPCC: high precision congestion control}.
\newblock In \bibinfo{booktitle}{\emph{Proceedings of the ACM Special Interest
  Group on Data Communication}}. \bibinfo{publisher}{ACM New York, NY, USA},
  \bibinfo{pages}{44--58}.
\newblock


\bibitem[\protect\citeauthoryear{Liang, Zhu, Jin, and Stoica}{Liang
  et~al\mbox{.}}{2019}]%
        {neurocuts}
\bibfield{author}{\bibinfo{person}{Eric Liang}, \bibinfo{person}{Hang Zhu},
  \bibinfo{person}{Xin Jin}, {and} \bibinfo{person}{Ion Stoica}.}
  \bibinfo{year}{2019}\natexlab{}.
\newblock \showarticletitle{Neural Packet Classification}. In
  \bibinfo{booktitle}{\emph{Proceedings of the ACM Special Interest Group on
  Data Communication}} (Beijing, China) \emph{(\bibinfo{series}{SIGCOMM '19})}.
  \bibinfo{publisher}{Association for Computing Machinery},
  \bibinfo{address}{New York, NY, USA}, \bibinfo{pages}{256–269}.
\newblock
\showISBNx{9781450359566}
\urldef\tempurl%
\url{https://doi.org/10.1145/3341302.3342221}
\showDOI{\tempurl}


\bibitem[\protect\citeauthoryear{Libraft}{Libraft}{[n.d.]}]%
        {libraft}
Libraft \bibinfo{year}{[n.d.]}\natexlab{}.
\newblock \bibinfo{title}{Libraft}.
\newblock \bibinfo{howpublished}{\url{https://github.com/willemt/raft}}.
\newblock
\newblock
\shownote{Accessed on 09/15/2020.}


\bibitem[\protect\citeauthoryear{Lim, Han, Andersen, and Kaminsky}{Lim
  et~al\mbox{.}}{2014}]%
        {mica}
\bibfield{author}{\bibinfo{person}{Hyeontaek Lim}, \bibinfo{person}{Dongsu
  Han}, \bibinfo{person}{David~G Andersen}, {and} \bibinfo{person}{Michael
  Kaminsky}.} \bibinfo{year}{2014}\natexlab{}.
\newblock \showarticletitle{$\{$MICA$\}$: A holistic approach to fast in-memory
  key-value storage}. In \bibinfo{booktitle}{\emph{11th $\{$USENIX$\}$
  Symposium on Networked Systems Design and Implementation ($\{$NSDI$\}$ 14)}}.
  \bibinfo{pages}{429--444}.
\newblock


\bibitem[\protect\citeauthoryear{Liu, Manousis, Vorsanger, Sekar, and
  Braverman}{Liu et~al\mbox{.}}{2016}]%
        {univmon}
\bibfield{author}{\bibinfo{person}{Zaoxing Liu}, \bibinfo{person}{Antonis
  Manousis}, \bibinfo{person}{Gregory Vorsanger}, \bibinfo{person}{Vyas Sekar},
  {and} \bibinfo{person}{Vladimir Braverman}.} \bibinfo{year}{2016}\natexlab{}.
\newblock \showarticletitle{One sketch to rule them all: Rethinking network
  flow monitoring with univmon}. In \bibinfo{booktitle}{\emph{Proceedings of
  the 2016 ACM SIGCOMM Conference}}. \bibinfo{pages}{101--114}.
\newblock


\bibitem[\protect\citeauthoryear{Miao, Zeng, Kim, Lee, and Yu}{Miao
  et~al\mbox{.}}{2017}]%
        {silkroad}
\bibfield{author}{\bibinfo{person}{Rui Miao}, \bibinfo{person}{Hongyi Zeng},
  \bibinfo{person}{Changhoon Kim}, \bibinfo{person}{Jeongkeun Lee}, {and}
  \bibinfo{person}{Minlan Yu}.} \bibinfo{year}{2017}\natexlab{}.
\newblock \showarticletitle{Silkroad: Making stateful layer-4 load balancing
  fast and cheap using switching asics}. In
  \bibinfo{booktitle}{\emph{Proceedings of the Conference of the ACM Special
  Interest Group on Data Communication}}. \bibinfo{pages}{15--28}.
\newblock


\bibitem[\protect\citeauthoryear{Narayana, Sivaraman, Nathan, Goyal, Arun,
  Alizadeh, Jeyakumar, and Kim}{Narayana et~al\mbox{.}}{2017}]%
        {marple}
\bibfield{author}{\bibinfo{person}{Srinivas Narayana}, \bibinfo{person}{Anirudh
  Sivaraman}, \bibinfo{person}{Vikram Nathan}, \bibinfo{person}{Prateesh
  Goyal}, \bibinfo{person}{Venkat Arun}, \bibinfo{person}{Mohammad Alizadeh},
  \bibinfo{person}{Vimalkumar Jeyakumar}, {and} \bibinfo{person}{Changhoon
  Kim}.} \bibinfo{year}{2017}\natexlab{}.
\newblock \showarticletitle{Language-directed hardware design for network
  performance monitoring}. In \bibinfo{booktitle}{\emph{Proceedings of the
  Conference of the ACM Special Interest Group on Data Communication}}.
  \bibinfo{pages}{85--98}.
\newblock


\bibitem[\protect\citeauthoryear{Ongaro and Ousterhout}{Ongaro and
  Ousterhout}{2014}]%
        {raft}
\bibfield{author}{\bibinfo{person}{Diego Ongaro} {and} \bibinfo{person}{John
  Ousterhout}.} \bibinfo{year}{2014}\natexlab{}.
\newblock \showarticletitle{In search of an understandable consensus
  algorithm}. In \bibinfo{booktitle}{\emph{2014 $\{$USENIX$\}$ Annual Technical
  Conference ($\{$USENIX$\}$$\{$ATC$\}$ 14)}}. \bibinfo{pages}{305--319}.
\newblock


\bibitem[\protect\citeauthoryear{ONOS}{ONOS}{[n.d.]}]%
        {onos}
ONOS \bibinfo{year}{[n.d.]}\natexlab{}.
\newblock \bibinfo{title}{ONOS}.
\newblock \bibinfo{howpublished}{\url{https://www.opennetworking.org/onos/}}.
\newblock
\newblock
\shownote{Accessed on 09/15/2020.}


\bibitem[\protect\citeauthoryear{Ousterhout, Gopalan, Gupta, Kejriwal, Lee,
  Montazeri, Ongaro, Park, Qin, Rosenblum, Rumble, Stutsman, and
  Yang}{Ousterhout et~al\mbox{.}}{2015}]%
        {ramcloud}
\bibfield{author}{\bibinfo{person}{John Ousterhout}, \bibinfo{person}{Arjun
  Gopalan}, \bibinfo{person}{Ashish Gupta}, \bibinfo{person}{Ankita Kejriwal},
  \bibinfo{person}{Collin Lee}, \bibinfo{person}{Behnam Montazeri},
  \bibinfo{person}{Diego Ongaro}, \bibinfo{person}{Seo~Jin Park},
  \bibinfo{person}{Henry Qin}, \bibinfo{person}{Mendel Rosenblum},
  \bibinfo{person}{Stephen Rumble}, \bibinfo{person}{Ryan Stutsman}, {and}
  \bibinfo{person}{Stephen Yang}.} \bibinfo{year}{2015}\natexlab{}.
\newblock \showarticletitle{The RAMCloud Storage System}.
\newblock \bibinfo{journal}{\emph{ACM Trans. Comput. Syst.}}
  \bibinfo{volume}{33}, \bibinfo{number}{3}, Article \bibinfo{articleno}{7}
  (\bibinfo{date}{Aug.} \bibinfo{year}{2015}), \bibinfo{numpages}{55}~pages.
\newblock
\showISSN{0734-2071}
\urldef\tempurl%
\url{https://doi.org/10.1145/2806887}
\showDOI{\tempurl}


\bibitem[\protect\citeauthoryear{Prekas, Kogias, and Bugnion}{Prekas
  et~al\mbox{.}}{2017}]%
        {zygos}
\bibfield{author}{\bibinfo{person}{George Prekas}, \bibinfo{person}{Marios
  Kogias}, {and} \bibinfo{person}{Edouard Bugnion}.}
  \bibinfo{year}{2017}\natexlab{}.
\newblock \showarticletitle{Zygos: Achieving low tail latency for
  microsecond-scale networked tasks}. In \bibinfo{booktitle}{\emph{Proceedings
  of the 26th Symposium on Operating Systems Principles}}.
  \bibinfo{pages}{325--341}.
\newblock


\bibitem[\protect\citeauthoryear{Rashelbach, Rottenstreich, and
  Silberstein}{Rashelbach et~al\mbox{.}}{2020}]%
        {nuevomatch}
\bibfield{author}{\bibinfo{person}{Alon Rashelbach}, \bibinfo{person}{Ori
  Rottenstreich}, {and} \bibinfo{person}{Mark Silberstein}.}
  \bibinfo{year}{2020}\natexlab{}.
\newblock \showarticletitle{A Computational Approach to Packet Classification}.
\newblock \bibinfo{journal}{\emph{arXiv preprint arXiv:2002.07584}}
  (\bibinfo{year}{2020}).
\newblock


\bibitem[\protect\citeauthoryear{Sutherland, Gupta, Falsafi, Marathe,
  Pnevmatikatos, and Daglis}{Sutherland et~al\mbox{.}}{2020}]%
        {nebula}
\bibfield{author}{\bibinfo{person}{Mark Sutherland}, \bibinfo{person}{Siddharth
  Gupta}, \bibinfo{person}{Babak Falsafi}, \bibinfo{person}{Virendra Marathe},
  \bibinfo{person}{Dionisios Pnevmatikatos}, {and} \bibinfo{person}{Alexandros
  Daglis}.} \bibinfo{year}{2020}\natexlab{}.
\newblock \bibinfo{booktitle}{\emph{The NEBULA RPC-Optimized Architecture}}.
\newblock \bibinfo{type}{{T}echnical {R}eport}. \bibinfo{institution}{EcoCloud,
  EPFL}.
\newblock


\bibitem[\protect\citeauthoryear{SX1036}{SX1036}{[n.d.]}]%
        {SX1036}
SX1036 \bibinfo{year}{[n.d.]}\natexlab{}.
\newblock \bibinfo{title}{SX1036 Product Brief}.
\newblock
  \bibinfo{howpublished}{\url{https://www.mellanox.com/related-docs/prod_eth_switches/PB_SX1036.pdf}}.
\newblock
\newblock
\shownote{Accessed on 09/15/2020.}


\bibitem[\protect\citeauthoryear{Taylor and Turner}{Taylor and Turner}{2007}]%
        {classbench}
\bibfield{author}{\bibinfo{person}{David~E. Taylor} {and}
  \bibinfo{person}{Jonathan~S. Turner}.} \bibinfo{year}{2007}\natexlab{}.
\newblock \showarticletitle{ClassBench: A Packet Classification Benchmark}.
\newblock \bibinfo{journal}{\emph{IEEE/ACM Trans. Netw.}} \bibinfo{volume}{15},
  \bibinfo{number}{3} (\bibinfo{date}{June} \bibinfo{year}{2007}),
  \bibinfo{pages}{499–511}.
\newblock
\showISSN{1063-6692}
\urldef\tempurl%
\url{https://doi.org/10.1109/TNET.2007.893156}
\showDOI{\tempurl}


\bibitem[\protect\citeauthoryear{Tofino}{Tofino}{[n.d.]}]%
        {tofino}
Tofino \bibinfo{year}{[n.d.]}\natexlab{}.
\newblock \bibinfo{title}{Tofino}.
\newblock
  \bibinfo{howpublished}{\url{https://www.barefootnetworks.com/products/brief-tofino/}}.
\newblock
\newblock
\shownote{Accessed on 02/04/2020.}


\bibitem[\protect\citeauthoryear{Van~Tu, Hyun, Kim, Yoo, and Hong}{Van~Tu
  et~al\mbox{.}}{2018}]%
        {intcollector}
\bibfield{author}{\bibinfo{person}{Nguyen Van~Tu}, \bibinfo{person}{Jonghwan
  Hyun}, \bibinfo{person}{Ga~Yeon Kim}, \bibinfo{person}{Jae-Hyoung Yoo}, {and}
  \bibinfo{person}{James Won-Ki Hong}.} \bibinfo{year}{2018}\natexlab{}.
\newblock \showarticletitle{Intcollector: A high-performance collector for
  in-band network telemetry}. In \bibinfo{booktitle}{\emph{2018 14th
  International Conference on Network and Service Management (CNSM)}}. IEEE,
  \bibinfo{pages}{10--18}.
\newblock


\bibitem[\protect\citeauthoryear{Verilator}{Verilator}{[n.d.]}]%
        {verilator}
Verilator \bibinfo{year}{[n.d.]}\natexlab{}.
\newblock \bibinfo{title}{Verilator}.
\newblock
  \bibinfo{howpublished}{\url{https://www.veripool.org/wiki/verilator}}.
\newblock
\newblock
\shownote{Accessed on 2020-01-29.}


\bibitem[\protect\citeauthoryear{Xilinx-TCAM}{Xilinx-TCAM}{[n.d.]}]%
        {xilinx-tcam}
Xilinx-TCAM \bibinfo{year}{[n.d.]}\natexlab{}.
\newblock \bibinfo{title}{Ternary Content Addressable Memory (TCAM) Search IP
  for SDNet}.
\newblock
  \bibinfo{howpublished}{\url{https://www.xilinx.com/support/documentation/ip_documentation/tcam/pg190-tcam.pdf}}.
\newblock
\newblock
\shownote{Accessed on 09/15/2020.}


\bibitem[\protect\citeauthoryear{Yaseen, Sonchack, and Liu}{Yaseen
  et~al\mbox{.}}{2018}]%
        {speedlight}
\bibfield{author}{\bibinfo{person}{Nofel Yaseen}, \bibinfo{person}{John
  Sonchack}, {and} \bibinfo{person}{Vincent Liu}.}
  \bibinfo{year}{2018}\natexlab{}.
\newblock \showarticletitle{Synchronized network snapshots}. In
  \bibinfo{booktitle}{\emph{Proceedings of the 2018 Conference of the ACM
  Special Interest Group on Data Communication}}. \bibinfo{pages}{402--416}.
\newblock


\bibitem[\protect\citeauthoryear{Yu, Sonchack, and Liu}{Yu
  et~al\mbox{.}}{2020}]%
        {mantis}
\bibfield{author}{\bibinfo{person}{Liangcheng Yu}, \bibinfo{person}{John
  Sonchack}, {and} \bibinfo{person}{Vincent Liu}.}
  \bibinfo{year}{2020}\natexlab{}.
\newblock \showarticletitle{Mantis: Reactive Programmable Switches}. In
  \bibinfo{booktitle}{\emph{Proceedings of the Annual conference of the ACM
  Special Interest Group on Data Communication on the applications,
  technologies, architectures, and protocols for computer communication}}.
  \bibinfo{pages}{296--309}.
\newblock


\bibitem[\protect\citeauthoryear{Zhang, Li, Wang, Liu, Chen, Hu, Gu, Li, Xu,
  and Wu}{Zhang et~al\mbox{.}}{2020}]%
        {poseidon}
\bibfield{author}{\bibinfo{person}{Menghao Zhang}, \bibinfo{person}{Guanyu Li},
  \bibinfo{person}{Shicheng Wang}, \bibinfo{person}{Chang Liu},
  \bibinfo{person}{Ang Chen}, \bibinfo{person}{Hongxin Hu},
  \bibinfo{person}{Guofei Gu}, \bibinfo{person}{Qianqian Li},
  \bibinfo{person}{Mingwei Xu}, {and} \bibinfo{person}{Jianping Wu}.}
  \bibinfo{year}{2020}\natexlab{}.
\newblock \showarticletitle{Poseidon: Mitigating volumetric ddos attacks with
  programmable switches}. In \bibinfo{booktitle}{\emph{Proceedings of NDSS}}.
\newblock


\bibitem[\protect\citeauthoryear{Zhang, Liu, Zeng, and Krishnamurthy}{Zhang
  et~al\mbox{.}}{2017}]%
        {microburst-measurements}
\bibfield{author}{\bibinfo{person}{Qiao Zhang}, \bibinfo{person}{Vincent Liu},
  \bibinfo{person}{Hongyi Zeng}, {and} \bibinfo{person}{Arvind Krishnamurthy}.}
  \bibinfo{year}{2017}\natexlab{}.
\newblock \showarticletitle{High-resolution measurement of data center
  microbursts}. In \bibinfo{booktitle}{\emph{Proceedings of the 2017 Internet
  Measurement Conference}}. \bibinfo{pages}{78--85}.
\newblock


\bibitem[\protect\citeauthoryear{Zhu, Eran, Firestone, Guo, Lipshteyn, Liron,
  Padhye, Raindel, Yahia, and Zhang}{Zhu et~al\mbox{.}}{2015}]%
        {dcqcn}
\bibfield{author}{\bibinfo{person}{Yibo Zhu}, \bibinfo{person}{Haggai Eran},
  \bibinfo{person}{Daniel Firestone}, \bibinfo{person}{Chuanxiong Guo},
  \bibinfo{person}{Marina Lipshteyn}, \bibinfo{person}{Yehonatan Liron},
  \bibinfo{person}{Jitendra Padhye}, \bibinfo{person}{Shachar Raindel},
  \bibinfo{person}{Mohamad~Haj Yahia}, {and} \bibinfo{person}{Ming Zhang}.}
  \bibinfo{year}{2015}\natexlab{}.
\newblock \showarticletitle{Congestion control for large-scale RDMA
  deployments}.
\newblock \bibinfo{journal}{\emph{ACM SIGCOMM Computer Communication Review}}
  \bibinfo{volume}{45}, \bibinfo{number}{4} (\bibinfo{year}{2015}),
  \bibinfo{pages}{523--536}.
\newblock


\end{thebibliography}

\label{totalpage}

\end{document}